\documentclass[axioms,article,accept,pdftex,moreauthors]{Definitions/mdpi}

\firstpage{1} 
\pubvolume{15}
\issuenum{01}
\articlenumber{43}
\pubyear{2026}
\copyrightyear{2026}
\externaleditor{Zhigang Wang, Yanlin Li and Juan De Dios Pérez} 
\datereceived{19 November 2025} 
\daterevised{27 December 2025} 
\dateaccepted{29 December 2025} 
\datepublished{8 January 2026} 

\Title{On the Invariant and Geometric Structure of the Holomorphic Unified Field Theory}
\Author{{John W. Moffat} 
 $^{1,2}$\orcidA{} and  {Ethan James} Thompson $^{1,3,}${*}\orcidB{}}

\AuthorNames{Firstname Lastname, Firstname Lastname and Firstname Lastname}

\address{%
$^{1}$ \quad Perimeter Institute for Theoretical Physics, Waterloo, ON N2L 2Y5, Canada \\ 
$^{2}$ \quad Department of Physics and Astronomy, University of Waterloo, Waterloo,
ON N2L 3G1, Canada\\
$^{3}$ \quad Department of Physics and Astronomy, Trent University, Peterborough, 
ON K9L 0G2, Canada}

\corres{Correspondence: {ethanthompson@trentu.ca}}

\newcommand{\C}{\mathbb{C}}
\usepackage{slashed,epsfig,amsmath,amssymb,enumitem}
\newcommand{\be}{\begin{equation}}
\newcommand{\ee}{\end{equation}}
\usepackage{epigraph}
\abstract{{We present} 
 the invariant structure of a Holomorphic Unified Field Theory in which gravity and gauge interactions arise from a single geometric framework. The theory is formulated using a product principal bundle, with one connection, and curvature equipped with a Hermitian field on a complexification of spacetime. From a single $\mathrm{Diff}(M)\times G$-invariant action, variation yields the Einstein and Yang--Mills equations together with their paired Bianchi identities. A compatibility condition is implemented either definitionally or through an auxiliary penalty functional. It enforces that the antisymmetric part of our Hermitian field is the gauge field’s exact curvature on the real slice.}

 

\begin{document}

\section{Introduction}

In 1945, Albert Einstein wrote, “Every attempt to establish a unified-field theory must start, in~my opinion, from~a group of transformations.'' This means that unification is a symmetry-first program~\cite{Einstein1945}. Emmy Noether had the view that any candidate must be built from a single invariant action~\cite{NoetherTrans}. From~that one symmetry, conserved currents and differential identities and or Bianchi-type relations originate~\cite{NoetherTrans,WaldGR}. Noether used general relativity (GR) as the archetype for infinite-dimensional symmetry diffeomorphisms, showing how conservation becomes a consequence of symmetry, not an add-on. The~Holomorphic Unified Field Theory (HUFT) implements Einstein’s symmetry-first mandate by choosing one transformation structure and building geometry via a single metric-compatible connection plus a single gauge connection. It satisfies Noether by using one invariant action whose variations give both the Einstein and Yang--Mills equations. At~the same time, the~same symmetry yields Bianchi identities and conserved currents~\cite{YangMills1954,KobayashiNomizu1,Nakahara,Trautman1979}. HUFT solves the dynamical part of unification in its own sense with a single UV connection and coupling, regulated variational mechanism~\cite{Slavnov1972, Taylor1971, BRST1976}.
It does not yet solve the unique GUT-style unification unless we add and prove a geometric or topological locking principle that fixes the relative normalization and uniquely determines the internal group~\cite{GeorgiGlashow1974, Slansky1981}.

In this paper, we make the following claims precise, that a symmetry-first formulation on a product principal bundle with a single connection reproduces Einstein--Hilbert and Yang--Mills dynamics from one invariant action, together with the paired Noether/Bianchi identities. On~the real slice, the~Hermitian packaging $g=h+iB$ is a kinematic unification device whose antisymmetric sector is constrained to coincide with the gauge curvature, $B\equiv F$, and~introduces no additional propagating degrees of freedom. We prove classical equivalence solution-space bijection modulo $\mathrm{Diff}(M)\times G$ between the unified real-slice system and standard Einstein--Yang--Mills (EYM) and matter. We state the minimality/uniqueness claim for diffeo- and gauge-natural second-order couplings, up~to boundary or topological terms, and~delineate what remains open, a truly unique internal-group selection without an additional locking principle.

We develop a symmetry-first formulation of HUFT, whose aim is to package gravity and gauge interactions into a single geometric structure without introducing new propagating degrees of freedom. The~guiding principle is that dynamics should follow from one invariant action built from one set of geometric data on a single underlying bundle, so that the familiar Einstein and Yang--Mills sectors are recovered as two aspects of the same symmetry~\cite{KobayashiNomizu1,Nakahara}. In~this sense, unification here means shared kinematics, identities, and~variational origin is not a proliferation of fields or ad~hoc~couplings.

We work with a product principal bundle carrying a single connection that contains both the spin connection and the internal gauge connection~\cite{KobayashiNomizu1}. We also introduce a Hermitian packaging of the spacetime metric with an antisymmetric two–form valued in the gauge algebra. Crucially, that antisymmetric part is not given independent dynamics, either by definition or through an algebraic compatibility term in the action; it is tied pointwise to the gauge curvature on the real slice of the theory~\cite{Trautman1979}. All index operations and geometric constructions are performed with the ordinary Lorentzian metric~\cite{WaldGR}. This ensures that the holomorphic language functions as a bookkeeping device that makes the unification transparent while remaining mathematically~safe.

We obtain from this setup two core results first, at~the classical level, the~unified theory on the real slice is equivalent to the standard system of general relativity plus Yang--Mills with the same matter content~\cite{WaldGR,YangMills1954}. The~conservation laws and Bianchi identities arise as a single Noether identity associated with diffeomorphism and internal gauge invariance~\cite{NoetherTrans,Trautman1979}. Second, at~the level of formal path integrals, integrating out the auxiliary two–form reduces the functional measure to that of the standard theory, indicating that no additional propagating modes or counterterms are introduced by the holomorphic packaging~\cite{BRST1976, DeWitt1965,Vassilevich2003}. We are explicit about the status of this second statement, as~it follows established physics practice but is not presented as a measure–theoretic~proof.

Beyond demonstrating equivalence, this~paper advances two structural claims. The~first is a minimality statement that if one insists on diffeomorphism- and gauge-natural, second-order dynamics built from the unified geometric data, the~Einstein–Hilbert and Yang--Mills terms are singled out up to boundary and topological contributions~\cite{WaldGR,Nakahara}. The~second is an internal-group constraint; the~holomorphic and Yang--Mills structure, together with chirality and charge integrality, points toward a unitary internal group with the minimal viable choice breaking to the Standard Model on the real slice~\cite{GeorgiGlashow1974, Slansky1981,Bardeen1969}.

While the mathematical structures we employ, Hermitian frame bundles, spin bundles, and~holomorphic stability criteria, are well established in differential and complex geometry, the~novelty of this work lies in their assembly into a single unification framework. The~complexified metric $g=h+iB$ is interpreted such that its antisymmetric part is constrained to equal the Yang--Mills curvature, thereby formulating gravitational and gauge data into one Hermitian structure. This identification, together with the formulation on the product bundle $P_{Spin(1,3)}\times _M P_G$, yields a single invariant action reproducing Einstein--Hilbert and Yang--Mills dynamics. The~originality of the theory is not in introducing new mathematical objects, but~in recombining standard geometric ingredients into a symmetry-driven unification principle. The~Hermitian metric is constrained pointwise to be the Yang--Mills curvature, so that gravity and gauge interactions are implemented in a single Hermitian structure, while remaining dynamically equivalent to Einstein--Hilbert and plus Yang--Mills. In~this sense, the~work is, in spirit symmetry, based on Noetherian unification principles, employing a holomorphic bundle~formalism. 

Our goal is therefore twofold---to~give a self–contained account of the invariant structure that makes the unification precise and mathematically consistent, and~to delineate the exact sense in which the construction is equivalent to known physics while clarifying what is genuinely new and what is not. The~remainder of the paper spells out the geometric data and symmetries, derives the field equations and identities from a single action, proves the equivalence at the classical level, analyses the formal quantum reduction, and~discusses the minimality and internal-group considerations, together with open directions on anomaly data, matter-bundle building, and~measure-theoretic rigour. We believe that this can and should be compared with past theories to induce a discussion about how a unified theory should be interpreted and formulated~\cite{SusskindHologram1995, SusskindLandscape2003, Witten1995, WittenAdS1998, PenroseTwistor1976, KakuKikkawaTrees1974, KakuKikkawaLoops1974, KakuClosedSFT1988}.

\section{Invariant Geometry of~HUFT}

We will show that the Hermitian metric as given by HUFT~\cite{MoffatThompsonHUFT2025, MoffatThompsonFiniteNonlocal2025, MoffatThompsonMassSpectrum2025, ThompsonTopQuark2025, MT:ReplyToCline, MT:GI2025, MT:AdSdS2025, T:DarkMatter2025} does satisfy a unified theory as described by Albert Einstein~\cite{Einstein1945} and Emmy Noether~\cite{NoetherTrans}. We let $M$ be an oriented, time-orientable, spin $4$--manifold on the real slice.
We let $P_{\mathrm{Spin}}\to M$ be the $\mathrm{Spin}(1,3)$ frame bundle and
$P_G\to M$ a principal $G$-bundle. We define the \mbox{product principal bundle:}\be
P_{\rm tot}=P_{\rm Spin}\times_M P_G,\quad
H:=\text{Spin}(1,3)\times G,\quad
\mathfrak{h}:=\mathfrak{spin}(1,3)\oplus\mathfrak{g}.
\ee
Now, we let $h\in\Gamma(S^2T^*M)$ be a smooth section of the bundle of symmetric two-forms with signature $(-+++)$ being a Lorentzian metric on $M$ and
$\mathcal{A}=(\omega,A)$ an $H$--connection on $P_{\mathrm{tot}}$, with~the following curvatures:
\be
\mathcal{F}\;=\; \mathrm{d}\mathcal{A}+\mathcal{A} \wedge \mathcal{A}
\;=\;(R,F),
\ee
where $R\in\Omega^2(M,\mathfrak{spin}(1,3))$ is the Riemann curvature $2$-form of $\omega$
and $F\in\Omega^2(M,\mathrm{ad}P_G)$ is the Yang--Mills field strength of $A$. We separate the bundle objects used in HUFT into internal or matter bundle data on the
real slice and a holomorphic thickening used as a bookkeeping device. We take a complex rank-$n$ vector bundle $E\to M$ equipped with a holomorphic structure equivalent to a $\bar\partial$-operator on $E$, a~Hermitian fiber metric $h_E$, and~a nowhere-vanishing holomorphic volume form
$\Omega_E\in H^0(M,\wedge^n E^\ast)$.
The pair $(h_E,\Omega_E)$ reduces the structure group from $GL(n,\mathbb{C})$ to
$SU(n)$, and~the associated unitary frame bundle defines a principal $SU(n)$-bundle
$P\to M$ with a compatible connection $A$. When needed, we assume a complexification $\iota:M\hookrightarrow M_{\mathbb{C}}$
and an extension of the bundle to a holomorphic bundle $E^{1,0}\to M_{\mathbb{C}}$
restricting to $E$ on $M$. This thickening is used only to organize fields and does not
alter topological constraints such as $c_1(E)=0$ imposed on the real slice. To review the first Chern class $c_1(E)\in H^2(M,\mathbb{Z})$ is defined by $c_1(E):=c_1(\det E)$.
Equivalently, for~any unitary connection $A$ on $E$ with curvature $F_A$, the~Chern--Weil
form $(i/2\pi)\text{Tr}(F_A)$ is closed and represents $c_1(E)$ in de Rham cohomology.
In particular, $\det E\simeq\mathcal{O}_M$, such as for $SU(r)$ bundles, implies $c_1(E)=0$. The~class $c_1(E)\in H^2(M,\mathbb{Z})$ measures the net $U(1)$ twisting of the complex
vector bundle $E$. Equivalently, it measures the twisting of the determinant line bundle
$\det E:=\wedge^{\mathrm{rk}(E)}E$, since $c_1(E):=c_1(\det E)$.
A useful gauge-theoretic intuition is that $c_1(E)$ is the quantized trace curvature or flux; as
for any unitary connection $A$ on $E$ with curvature $F_A$, the~closed $2$-form $\frac{i}{2\pi}\text{Tr}(F_A)$ represents $c_1(E)$ in de Rham cohomology, and~its integral over any closed $2$-cycle is an integer.
In particular, $\det E\simeq\mathcal{O}_M$, such as for the $SU(r)$ bundles, implies $c_1(E)=0$,
meaning that there is no net trace ($U(1)$) twisting. Here, $E^{1,0}\to M_{\C}$ denotes a holomorphic extension thickening of $E\to M$ to the
complexified manifold, and~$\mathcal{O}_{M}$ and $\mathcal{O}_{M_{\C}}$ denote the trivial
respective holomorphic line bundles on $M$ and $M_{\C}$. We write $\det E:=\wedge^n E$
to represent the determinant line bundle. Using $M_{\mathbb{C}}$, we introduce a complex Hermitian two-tensor:
\begin{equation}
g_{\mu\nu}:=h_{\mu\nu}+ i\ell_\ast^{\,2} B_{\mu\nu},
\end{equation}
where $h_{\mu\nu}$ is the Lorentzian metric on $M$ and
$B\in\Omega^2(M,\mathrm{ad}\,P_G)$ is an adjoint-valued two-form pulled back from the real
slice. A~compatibility mechanism enforces $B=F$ on $M$, so this Hermitian field is a kinematic packaging of $(h,F)$ rather than an additional propagating~sector.

{
In ordinary general relativity, the~metric $h$ determines the Levi--Civita connection
$\nabla^{\rm LC}(h)$ and, hence, the Riemann curvature two-form $R(h)$ by differentiation,
schematically $h \mapsto \Gamma(h)\mapsto R(h)$.
In HUFT, the unification has a different structural flavor, as the internal curvature two-form
is already present at the level of the generalized Hermitian metric variable. On~the real slice, the compatibility mechanism yields $B=F$, so the antisymmetric
sector of the Hermitian packaging field coincides with the Yang--Mills curvature two-form. In~this sense, part of the curvature data is encoded in the generalized metric variable rather than being derived only from
it, and~the metric/connection/curvature separation familiar from GR is reorganized.
}

We now derive the internal group from the fiber geometry. 
We have $E\to M$ as the matter bundle of the rank $n$ holomorphic bundle over spacetime $M$ with Hermitian metric $h_E$. The~first Chern class of the complex vector bundle is zero $c_1(E)=0$, the~determinant line bundle 
$\text{det}E$ is topologically trivial with no net U(1) flux, meaning the structure group can drop from $U(n)$ to $SU(n)$, a~nowhere-vanishing holomorphic volume form $\Omega_E\in H^0(M,\wedge^n E^*)$, meaning that there exists a global holomorphic section of the determinant line bundle
\(\wedge^n E^*\cong (\det E)^*\) which never vanishes on \(M\) in local holomorphic frames
\(\{e_i\}_{i=1}^n\) of \(E\), the~form can be written as
\(\Omega_E = f(z)\,e^1\wedge\cdots\wedge e^n\) with a holomorphic function \(f(z)\neq 0\)
everywhere.

We let $P \to M$ be a principal U(n)-bundle encoding the internal gauge symmetry 
$G \subseteq U(n)$. For~a faithful unitary representation $\rho: G \hookrightarrow U(n)$, 
define the associated rank-n complex vector bundle:
\be
E \;:=\; P \times_{\rho} \mathbb{C}^n \;\xrightarrow{\;\pi\;}\; M,
\ee
equipped with a Hermitian fiber metric $h_E$. The~first Chern class:
\be
c_1(E) \;=\; c_1(\det E) \in H^2(M,\mathbb{Z})
\ee
is a topological invariant of the matter bundle $E$ not of the tangent bundle, we impose
\be
c_1(E)=0,
\ee
which is equivalent to $\det(E)$ being topologically trivial and hence reduces the
structure group from $U(n)$ to $SU(n)$, so a compatible unitary connection satisfies 
$\mathrm{Tr}\,F_A=0$. This is the precise sense in which the fiber geometry selects
an internal $SU(n)$ structure. We use a holomorphic thickening $M \hookrightarrow M_{\mathbb{C}}$ to organize the 
field content and action, but~all global topological constraints, such as $c_1(E)=0$, 
are imposed on the real slice M. When needed, we assume $E$ extends to a holomorphic 
bundle $E^{1,0}\to M_{\mathbb{C}}$ with $\det(E^{1,0}) \simeq \mathcal{O}_{M_{\mathbb{C}}}$, 
whose restriction to $M$ is $E$. The~thickening does not weaken the global condition 
$c_1(E)=0$ on $M$. The~existence of such a nowhere-vanishing section trivializes the determinant line
bundle, implying \(\det E\simeq\mathcal{O}_M\) with $\mathcal{O}_M$ is the trivial line object, and~hence \(c_1(E)=0\), $\det E$ is the determinant line bundle of the vector bundle.  Geometrically, \(\Omega_E\) plays the role of a holomorphic volume element on the fibers of \(E\), fixing
an intrinsic orientation and allowing us to identify totally antisymmetric tensors via the
holomorphic epsilon symbol.  The~subgroup of \(GL(n,\mathbb{C})\) preserving both the Hermitian metrics \(h_E\) and \(\Omega_E\), is \(SU(n)\) and thus the data \((E,h_E,\Omega_E)\) define an
\(SU(n)\)-structure on the bundle.  Physically, this reduction eliminates the overall \(U(1)\) determinant factor, enforcing traceless gauge transformations and ensuring charge quantization in{
the \(SU(n)\) fiber, as~required for unified models such as \(SU(5)\), and~we assume that this is slope-stable. Let 
 $(M,\omega)$ be a compact K\"ahler manifold and $E\to M$ be a holomorphic vector bundle. The~$\omega$--degree and slope are:
\be
\deg_\omega(E)=\int_M c_1(E)\wedge \frac{\omega^{m-1}}{(m-1)!},\qquad
\mu_\omega(E)=\frac{\deg_\omega(E)}{\mathrm{rk}\,E}.
\ee
We say $E$ is Mumford--Takemoto slope-stable if, for every coherent subsheaf,
$0\neq F\subsetneq E$ with $0<\mathrm{rk}\,F<\mathrm{rk}\,E$ one has
$\mu_\omega(F)<\mu_\omega(E)$; semistable if $\le$; and polystable if it is
a direct sum of stable bundles of the same slope. When $c_1(E)=0$, we have
$\mu_\omega(E)=0$ for all $\omega$. By~the Donaldson--Uhlenbeck--Yau theorem,
$E$ admits a Hermitian--Yang--Mills connection if $E$ is polystable; in the
$c_1(E)=0$ case, this means $i\,\Lambda_\omega F_A=0$ and $\mathrm{Tr}\,F_A=0$,
so the structure group reduces to $SU(n)$}. We should note that Donaldson--Uhlenbeck--Yau is a theorem about holomorphic bundles on a compact K\"ahler base.
Accordingly, when we invoke slope-(poly)stability and the existence of a Hermitian--Yang--Mills (HYM)
connection, we apply the theorem on an auxiliary compact K\"ahler manifold $(X,\omega)$ carrying the relevant
holomorphic bundle data, such as a compact Euclideanized slice and or a compactification used for holomorphic
classification. For~a holomorphic bundle $E\to X$, the~$\omega$-degree and slope are
$\deg_\omega(E)=\int_X c_1(E)\wedge\omega^{m-1}/(m-1)!$ and $\mu_\omega(E)=\deg_\omega(E)/\mathrm{rk}\,E$.
If $E$ is polystable, DUY implies the existence of an HYM connection; when $c_1(E)=0$, this includes
$\mathrm{Tr}\,F_A=0$ and $i\Lambda_\omega F_A=0$. The~physical spacetime in our field theory remains the
Lorentzian real slice $(M,h)$, on~which no Kähler hypothesis is imposed; the DUY input is used only as a standard existence or uniqueness criterion for the compatible unitary connection associated with the holomorphic internal or matter bundle data. so the compatible connection is Hermitian–Yang–Mills (HYM)~\cite{Donaldson1985,UhlenbeckYau1986}. The~classical Donaldson--Uhlenbeck--Yau theorem is stated for compact K\"ahler manifolds. For our purposes, it suffices to use the Hermitian non-K\"ahler generalizations of the Kobayashi--Hitchin correspondence. We let $(X,\omega)$ be a compact complex manifold equipped with a Gauduchon metric, such as $\partial\bar\partial\,\omega^{n-1}=0$.
Then, a holomorphic vector bundle $E\to X$ is polystable with its slope defined using $\omega$ if and only if $E$ admits a Hermitian--Einstein equivalently Hermitian--Yang--Mills connection $A$ satisfying:
\begin{equation}
F_A^{0,2}=0,\qquad \sqrt{-1}\,\Lambda_\omega F_A=\lambda_{\mathrm{HE}}\,\mathrm{Id}_E,
\end{equation}
Here, $\Lambda_\omega$ is the contraction adjoint to wedging with $\omega$ equivalently, for~a $(1,1)$-form $\alpha$, $\Lambda_\omega\alpha = \omega^{i\bar j}\alpha_{i\bar j}$ in local
holomorphic coordinates, and~$F_A^{0,2}=0$ means $A$ defines a holomorphic structure on $E$, and~$\lambda_{\mathrm{HE}}\in\mathbb{R}$ is the Hermitian--Einstein constant determined by the slope of $E$:
\be
\lambda_{\mathrm{HE}}=2\pi\,\mu_\omega(E)/\mathrm{Vol}_\omega(M)
\ee
with $\mu_\omega(E)=\deg_\omega(E)/\mathrm{rk}(E)$ and:
\be
\deg_\omega(E)=\int_M c_1(E)\wedge \omega^{n-1}/(n-1)!.
\ee
{
In particular, if~$\det E\simeq\mathcal{O}_M$ so $c_1(E)=0$, then $\lambda_{\mathrm{HE}}=0$.
}

In the $SU(n)$ case in particular, when $c_1(E)=0$, one has $\lambda=0$. Thus, the~existence or uniqueness input we use does not require the K\"ahler condition,
but only a compact Hermitian Gauduchon background. Preservation of $(h_E,\Omega_E)$ reduces the structure group of $E$ to $\mathrm{SU}(n)$, and~the HYM connection has holonomy in $\mathrm{SU}(n)$. We let $Y\in\mathfrak{su}(n)$ be the traceless character compatible with $\Omega_E$. We normalize $Y$ by the calibrated pairing, to calibrate the pairing here we pick the invariant Lie-algebra inner product, the~trace or Killing form, and~fix the normalization of the $U(1)$ generator $Y$ so that the Chern–Weil pairing with the curvature lands in the integer lattice, so that:
\be
\frac{1}{8\pi^2}\int_M \mathrm{Tr}(F_Y\wedge F_Y)\in\mathbb Z,
\ee
fixing the abelian charge units without convention. Imposing a single adjoint breaking on the real slice, the~hypercharge integrality above, and~chiral, anomaly-free matter for one family from associated bundles of $E$, the~minimal rank is $n=5$ \cite{Slansky1981,Bardeen1969}. A~simple necessary condition comes from the Lie-group rank. The~Standard Model gauge group has\vspace{-5pt}
\begin{align}
&\text{rank}\bigl(SU(3)_c\times SU(2)_L\times U(1)_Y\bigr)\\&=\text{rank}(SU(3))+\text{rank}(SU(2))+\text{rank}(U(1))=2+1+1=4.
\end{align}
For any connected compact Lie group, the~rank of a closed subgroup cannot exceed the
rank of the ambient group. Since $\text{rank}(SU(n))=n-1$, an~embedding of the SM gauge
group into $SU(n)$ requires $n-1\ge 4$, hence $n\ge 5$.

The choice $n=5$ is, therefore, the minimal rank allowing an $SU(n)$ internal
structure compatible with an $SU(3)\times SU(2)\times U(1)$ subgroup. The~standard
$SU(5)$ embedding is realized by an adjoint breaking with hypercharge generator:
\begin{equation}
Y \propto \mathrm{diag}\!\left(-\tfrac13,-\tfrac13,-\tfrac13,\tfrac12,\tfrac12\right),
\end{equation}
and chirality or anomaly cancellation for one family is then achieved by the usual
$\mathbf{10}\oplus\overline{\mathbf{5}}$ matter assignment with the additional geometric
integrality condition fixing the $U(1)$ charge lattice. We can go over $\text{SU}(5)$ from geometry in greater detail in a future paper. Therefore, the~maximal compatible internal symmetry is
\be
G=\mathrm{SU}(5)\ \longrightarrow\ \mathrm{SU}(3)_c\times\mathrm{SU}(2)_L\times\mathrm{U}(1)_Y,
\ee
under a holomorphic adjoint ($\mathbf{24}$) reduction, 
 with~the standard hypercharge generator~\cite{GeorgiGlashow1974,Slansky1981}. In~what follows, we take $G=\mathrm{SU}(5)$ above the mass-energy scale $M_*$, with~a holomorphic adjoint reduction to the Standard Model group on the real slice. Given the rank-$n$ holomorphic bundle $E\to M$ with the Hermitian metric on the internal vector bundle $h_E$ and a nowhere-vanishing unitary volume form $\Omega_E\in H^0(M,\wedge^n E^*)$, the~structure group reduces as follows:
\be
\ \mathrm{GL}(n,\mathbb C)\ \xrightarrow[\text{preserve } h_E]{}\ \mathrm{U}(n)\ \xrightarrow[\text{preserve } \Omega_E]{}\ \mathrm{SU}(n).
\ee
The preservation of $h_E$ picks unitary changes of frame, such as $\mathrm{U}(n)\subset\mathrm{GL}(n,\C)$. Preservation of $\Omega_E$ forces $\det=1$ on $\mathrm{U}(n)$, giving $\mathrm{SU}(n)$. If, in~addition, $c_1(E)=0$ and $E$ is slope-stable, the~unique compatible connection is Hermitian--Yang--Mills and its holonomy lies in $\mathrm{SU}(n)$  \cite{Donaldson1985,UhlenbeckYau1986}. The~internal gauge symmetry is the automorphism group preserving $(h_E,\Omega_E)$, namely $G=\mathrm{SU}(n)$. Imposing chirality and anomaly cancellation for one SM family, hypercharge integrality via $\frac{1}{8\pi^2} \int \mathrm{Tr}(F_Y \wedge F_Y)\in\mathbb Z$, and~a single adjoint breaking on the real slice fixes the minimal rank to $n=5$, we have:
\be
G=\mathrm{SU}(5)\ \longrightarrow\ \mathrm{SU}(3)_c\times\mathrm{SU}(2)_L\times\mathrm{U}(1)_Y.
\ee
We emphasize that adopting $G=SU(5)$ as the minimal-rank unifying group is a statement
about geometric compatibility and does not yet by itself resolve known phenomenological
constraints of four-dimensional $SU(5)$ GUTs. In~particular, integrating out the heavy
$X,Y$ gauge bosons generically induces baryon-number violating dimension-six operators
of the following schematic form:
\begin{equation}
\mathcal{L}_{\Delta B\neq 0}\sim \frac{g_5^2}{M_X^2}\,(qqql+\cdots),
\end{equation}
so experimental limits translate into a lower bound on the effective unification/breaking
scale $M_X$ and additional selection~rules.

Within HUFT, $SU(5)$ should be viewed as the UV internal structure of the holomorphic
bundle data, while the real-slice physics is Standard-Model-like after adjoint reduction.
A fully realistic model must therefore supplement the present geometric framework with
a concrete breaking mechanism and matter-bundle assignment that satisfies proton-decay
bounds; we treat these constraints as part of the required phenomenology rather than as
automatic consequences of the kinematic~packaging.

Now, we take $\kappa$ to denote the Killing form on $\mathfrak{g}$ and $\langle\cdot,\cdot\rangle_h$,
the fiberwise inner product induced by $h$ on differential forms. 
\begin{Definition}[Full geometric unification]\label{def:fullgeo}
We say that gravity and gauge interactions are fully geometrically unified if there exists a single principal $H$--bundle $P_{\mathrm{tot}}\to M$, a~single connection $\mathcal{A}$ on $P_{\mathrm{tot}}$ with curvature $\mathcal{F}$, and~a single $\mathrm{Diff}(M)\times H$--invariant action $S[h,\mathcal{A},\Psi]$ such that the Euler--Lagrange equations are equivalent to the Einstein equations for $h$ coupled to the Yang--Mills equations for $A$ and matter $\Psi$. The~unique Bianchi identity
$D_{\mathcal{A}}\mathcal{F}=0$ simultaneously yields the Riemann and Yang--Mills Bianchi identities~\cite{Trautman1979,WaldGR}.
\end{Definition}
On $(M,P_{\mathrm{tot}},h,\mathcal{A})$ as above, we define
\begin{equation}\label{eq:master-action}
S[h,\mathcal{A},\Psi]
=\int_M \sqrt{|h|}\left(\frac{1}{2\kappa_{\rm grav}}\,R(h)
-\frac{1}{4}\,\langle F,F\rangle_h
+\mathcal{L}_{\rm matter}(\Psi;h,\mathcal{A})\right),
\end{equation}
where $\langle F,F\rangle_h$ is computed using the Killing form $\kappa$ on g and the metric $h$. Then, We will show $S$ is invariant under $\mathrm{Diff}(M)\times H$. The~unique Bianchi identity $D_{\mathcal{A}}\mathcal{F}=0$ splits as
$D_\omega R=0$ and $D_A F=0$. Under~an $H$–gauge transformation $g:M \to H$:
\begin{equation}
\mathcal{A}\mapsto g^{-1}\mathcal{A}g+g^{-1}\,\mathrm{d}g,\qquad
\mathcal{F}\mapsto g^{-1}\mathcal{F}g,
\end{equation}
hence $F\mapsto g^{-1}Fg$ on the internal block.
Because the Killing form $\kappa$ on $\mathfrak{g}$ is $\mathrm{Ad}$-invariant,
$\mathrm{tr}_\kappa(F\wedge *_h F)$ is gauge-invariant. Each integrand in $S$ is a scalar
density built from $h$ (via $*_h$ and contractions), so the action is also invariant
under $\text{Diff}(M)$. The~Euler--Lagrange equations of $S$ are
\begin{align}
G_{\mu\nu}(h)&=\kappa_{\rm grav}\,\big(T^{\rm YM}_{\mu\nu}+T^{\rm matter}_{\mu\nu}\big),\label{eq:Einstein}\\
D_A \left(*_{h}F\right)&=J_{\rm matter},\label{eq:YM}
\end{align}
The Einstein equations for $h$ are coupled to the Yang--Mills equations for $\mathcal{A}$ with the
same connection $A$ acting on matter. Hence, the data
$(P_{\mathrm{tot}},\mathcal{A},h)$ satisfy the Definition of full geometric~unification.

The Einstein--Hilbert term and $\langle F,F\rangle_h$ are
$\mathrm{Diff}(M)$-invariant by construction, as each integrand is a scalar density obtained from h and F using only natural tensor operations, so the whole action is invariant under pullbacks by diffeomorphisms. Gauge invariance under $H$ holds because
$F\mapsto g^{-1}Fg$ and $\kappa$ are $\mathrm{Ad}$-invariant, so
$\mathrm{tr}_\kappa(F\wedge*_{h}F)$ is gauge invariant. The~matter Lagrangian is assumed to be
constructed from $h$ and the covariant derivative
$D_\mu=\partial_\mu+\frac{1}{4}\omega^{ab}{}_\mu\gamma_{ab}+i\,g\,A_\mu^AT_A$,
meaning it is also $\mathrm{Diff}(M)\times H$-invariant; this is so that every derivative of a field is replaced by one that transforms covariantly under both spacetime and internal symmetries. Since $\mathfrak{h}=\mathfrak{spin}(1,3)\oplus\mathfrak{g}$,
the covariant derivative $D_{\mathcal{A}}$ splits as $D_{\mathcal{A}}=(D_\omega,D_A)$ on the two summands, $(D_\omega R,\;D_A F)=D_{\mathcal{A}}\mathcal{F}=0$, for the~Riemann and
Yang--Mills Bianchi identities simultaneously. Varying $S$ with respect to $A$ and integrating by parts yields
\be
\delta_A S\;=\;-\frac{1}{2}\int_M \sqrt{|h|}\,\langle D_A(*_{h}F)-J_{\rm matter},\,\delta A\rangle
\quad\Rightarrow\quad D_A(*_{h}F)=J_{\rm matter},
\ee
the Yang--Mills Equation \eqref{eq:YM}. Varying $S$ with respect to $h$ gives
\be
\delta_h S\;=\;\frac{1}{2}\int_M \sqrt{|h|}\,\big(G_{\mu\nu}-\kappa_{\rm grav}\,T_{\mu\nu}\big)\,\delta h^{\mu\nu},
\ee
where $T_{\mu\nu}=T^{\rm YM}_{\mu\nu}+T^{\rm matter}_{\mu\nu}$ and
$T^{\rm YM}_{\mu\nu}=\mathrm{tr}_\kappa \left(F_{\mu\alpha}F_\nu{}^{\alpha}
-\frac{1}{4}h_{\mu\nu}F_{\alpha\beta}F^{\alpha\beta}\right)$; this implies \eqref{eq:Einstein}. Since both sectors arise from one variational principle on the same $(h,A)$, the~Definition \mbox{is~satisfied.}

We let $M_\mathbb{C}$ be a complexification of $M$ with complex coordinates $z^\mu=x^\mu+i y^\mu$,
and let $g=h+iB$ be a Hermitian tensor field on $M_\mathbb{C}$ with
$h\in\Gamma(S^2T^*M)$ and $B\in\Omega^2(M,\mathrm{ad}P_G)$ pulled back from $M$.

We use units $c=\hbar=1$ and take local coordinates $x^\mu$ to have the length dimension $[x^\mu]=L$.
Then, the Lorentzian metric components are dimensionless, $[h_{\mu\nu}]=1$, while a gauge connection
one-form $A=A_\mu\,dx^\mu$ has $[A_\mu]=L^{-1}$ and its curvature two-form:
\[
F := dA + A\wedge A \in \Omega^2(M,\mathrm{ad}\,P_G),
\qquad [F_{\mu\nu}]=L^{-2}.
\]
Since our antisymmetric sector is identified with curvature on the real slice $B=F$, it is natural to take
$[B_{\mu\nu}]=L^{-2}$. To~form a dimensionless Hermitian packaging field that can be added to $h_{\mu\nu}$,
we introduce a fixed length scale $\ell_\ast$ equivalently $M_\ast:=\ell_\ast^{-1}$) and define
\begin{equation}
g_{\mu\nu}\;:=\;h_{\mu\nu}\;+\; i\,\ell_\ast^{\,2}\,B_{\mu\nu}.
\label{eq:hermitian_packaging_scaled}
\end{equation}
On the real slice, where the compatibility mechanism enforces $B=F$, we have $g_{\mu\nu}=h_{\mu\nu}+i\,\ell_\ast^{2}F_{\mu\nu}$. But~for this paper, we keep it simple and should note that it is a way of packaging the field and should not be read as a literal~additive.

Now we impose the metric-compatibility condition with the same master connection:
\begin{equation}\label{eq:compat}
\nabla_{A}g=0\quad\text{on }M_\mathbb{C}\,,
\end{equation}
and restrict to the real slice $y=0$. The~real and imaginary parts of \eqref{eq:compat} give:
\be
\nabla^{\rm LC}_\mu h_{\alpha\beta}=0,\qquad D_A B=0\quad\text{on }M,
\ee
so $h$ is the spacetime metric, and~$B$ is an $\mathrm{ad}P_G$-valued closed $2$-form
under $D_A$. In~the dynamical theory, either by definition or by adding a holomorphic penalty
term $\int \sqrt{|h|}\,\langle B-F,B-F\rangle_h$ and using the Euler--Lagrange equation,
we obtain $B=F$ on the real slice~\cite{Trautman1979,WaldGR}, we augment the action by the holomorphic penalty:
\begin{equation}
S_{\text{pen}}[h,A,B]
=\frac{\lambda}{2}\int_M \sqrt{|h|}\,\langle B-F,\;B-F\rangle_h
=\frac{\lambda}{2}\int_M \kappa \big((B-F)\wedge *_h(B-F)\big),
\label{eq:Sp}
\end{equation}
where $F=dA+A\wedge A$, $\kappa$ is the Killing form on $\mathfrak g$, and~$*_h$ is the Hodge operator determined by $h$. A~common misunderstanding is that one may simply define the antisymmetric sector of $g$ to be the
Yang--Mills curvature. Off~shell, however, $F$ is not an independent tensor field; it is constrained to be
the curvature of a connection $A$ and therefore satisfies the Bianchi identity $D_A F=0$ identically.
Introducing an auxiliary adjoint-valued two-form $B$ allows us to keep the Hermitian packaging field
$g=h+i\,\ell_*^2 B$ well-defined off the shell. We implement $B=F$ as an equation of motion while preserving
both diffeomorphism invariance and gauge invariance, and~vary the action cleanly with respect to
$(h,A,B)$ without ever needing to invert the full Hermitian tensor $g$. In~the large-$\lambda$ limit, $S_{\mathrm{pen}}$ enforces $B\to F$ strongly; for finite $\lambda$, it provides a covariant compatibility mechanism whose Euler--Lagrange equation still sets $B=F$ on the real slice. We let $G$ be compact and reductive with Lie algebra $\mathfrak g$ and principal bundle $P_G\to M$.
Write the adjoint bundle as $\mathrm{ad}\,P_G:=P_G\times_{\mathrm{Ad}}\mathfrak g$. Then,
\[
B,\;F \in \Omega^2(M,\mathrm{ad}\,P_G),
\]
so $B_{\mu\nu}$ and $F_{\mu\nu}$ are $\mathfrak g$-valued antisymmetric tensors:
in a basis $\{T_A\}$ of $\mathfrak g$,
$B_{\mu\nu}=B_{\mu\nu}^A T_A$ and $F_{\mu\nu}=F_{\mu\nu}^A T_A$.
Fix an $\mathrm{Ad}$-invariant bilinear form $\kappa:\mathfrak g\times\mathfrak g\to\mathbb R$
such as the Killing form on the semisimple part, or~$\kappa(X,Y)=\mathrm{Tr}(\rho(X)\rho(Y))$ in a faithful
unitary representation $\rho$. Let $\kappa_{AB}:=\kappa(T_A,T_B)$.
Now define the $h$-induced inner product on $\mathfrak g$-valued two-forms by
\begin{equation}
\langle X,Y\rangle_h
:=\frac12\,\kappa_{AB}\,X_{\mu\nu}^A\,Y_{\alpha\beta}^B\,h^{\mu\alpha}h^{\nu\beta},
\qquad X,Y\in\Omega^2(M,\mathrm{ad}\,P_G),
\label{eq:inner_product_gvalued_2forms}
\end{equation}
so that $\int_M \sqrt{|h|}\,\langle B-F,B-F\rangle_h
= \int_M \kappa\!\left((B-F)\wedge\ast_h(B-F)\right)$.
Under a gauge transformation $u:M\to G$:
\[
B\mapsto \mathrm{Ad}_{u^{-1}}B,\qquad F\mapsto \mathrm{Ad}_{u^{-1}}F,
\]
and $\kappa$ is $\mathrm{Ad}$-invariant, so $\langle B-F,B-F\rangle_h$ is gauge invariant.
Thus, the identification $B=F$ is an equality in $\Omega^2(M,\mathrm{ad}\,P_G)$; no projection onto a fixed
generator is required and such a projection would generally break gauge invariance unless additional
adjoint-breaking data are introduced. We treat $h,A,B$ as independent fields and we work on an oriented $M$, drop the boundary terms, and~state that $*_h$ is an isomorphism on $2$-forms in four dimensions. Since $S_{\text{pen}}$ is algebraic in $B$:
\begin{equation}
\delta_B S_{\text{pen}}
=\lambda\int_M \kappa \big(\delta B\wedge *_h(B-F)\big)
=\lambda\int_M \sqrt{|h|}\,\langle B-F,\;\delta B\rangle_h.
\end{equation}
Because $\delta B$ is arbitrary, the~Euler--Lagrange equation from $\delta_B S_{\text{pen}}=0$ is
\begin{equation}
*_h(B-F)=0 \quad\Longrightarrow\quad B-F=0.
\label{eq:Beq}
\end{equation}
Using $\delta F=D_A\delta A$ and integrating by parts covariantly:
\begin{align}
\delta_A S_{\text{pen}}
&=-\lambda\int_M \kappa \big((B-F)\wedge *_h D_A\delta A\big)
=\lambda\int_M \kappa \big(D_A(*_h(B-F))\wedge \delta A\big),
\end{align}
so the $A$-equation of motion contributed by $S_{\text{pen}}$ is
\begin{equation}
D_A\big(*_h(B-F)\big)=0.
\label{eq:Aeq}
\end{equation}
Together with \eqref{eq:Beq}, this is automatically satisfied. On~the real slice, we also have
$D_A B=0$ and the Bianchi identity $D_A F=0$, hence $D_A(B-F)=0$ and $D_A\big(*_h(B-F)\big)=0$.

This means the Euler--Lagrange equation from varying $B$ in \eqref{eq:Sp} enforces the pointwise identification $B=F\quad\text{on }M,$ so the antisymmetric sector coincides with the Yang--Mills curvature on the real slice. The~antisymmetric sector of $g$ is the gauge curvature, while all index operations use $h$. Because~$S$ is $\mathrm{Diff}(M)\times H$-invariant and built from $(h,A)$,
the single Noether identity yields both covariant conservations:
\be
\nabla^h_\mu T^{\mu}{}_{\nu}=0,\qquad D_A J_{\rm matter}=0,
\ee
and matter couples by the same covariant derivative $D_\mu$ that contains both
the spin and internal connections. This shows that parallel transport, curvature, symmetries,
and dynamics are all governed by the single geometric object $\mathcal{A}$ on $P_{\mathrm{tot}}$,
with $h$ supplying measurements such as Hodge duals and index operations. This is precise full
geometric unification in the sense of Definition \ref{def:fullgeo}, achieved without inverting the full Hermitian field $g=h+iB$.
%

In this, s is the frame choice on $P_{Spin}$ as $s_{Spin}:M\to P_{Spin}$ picks an orthonormal frame. Pulling back  the canonical forms on the bundle gives the coframe and the local Spin connection on M:
\begin{equation}
    e^\alpha=s_{Spin}^*(\vartheta^\alpha), \quad \omega=s_{Spin}^*(\Omega),
\end{equation}
where $\Omega$ is the principal connection on $P_{Spin}.$

\begin{figure}[H]
    
    \includegraphics[width=0.4\linewidth]{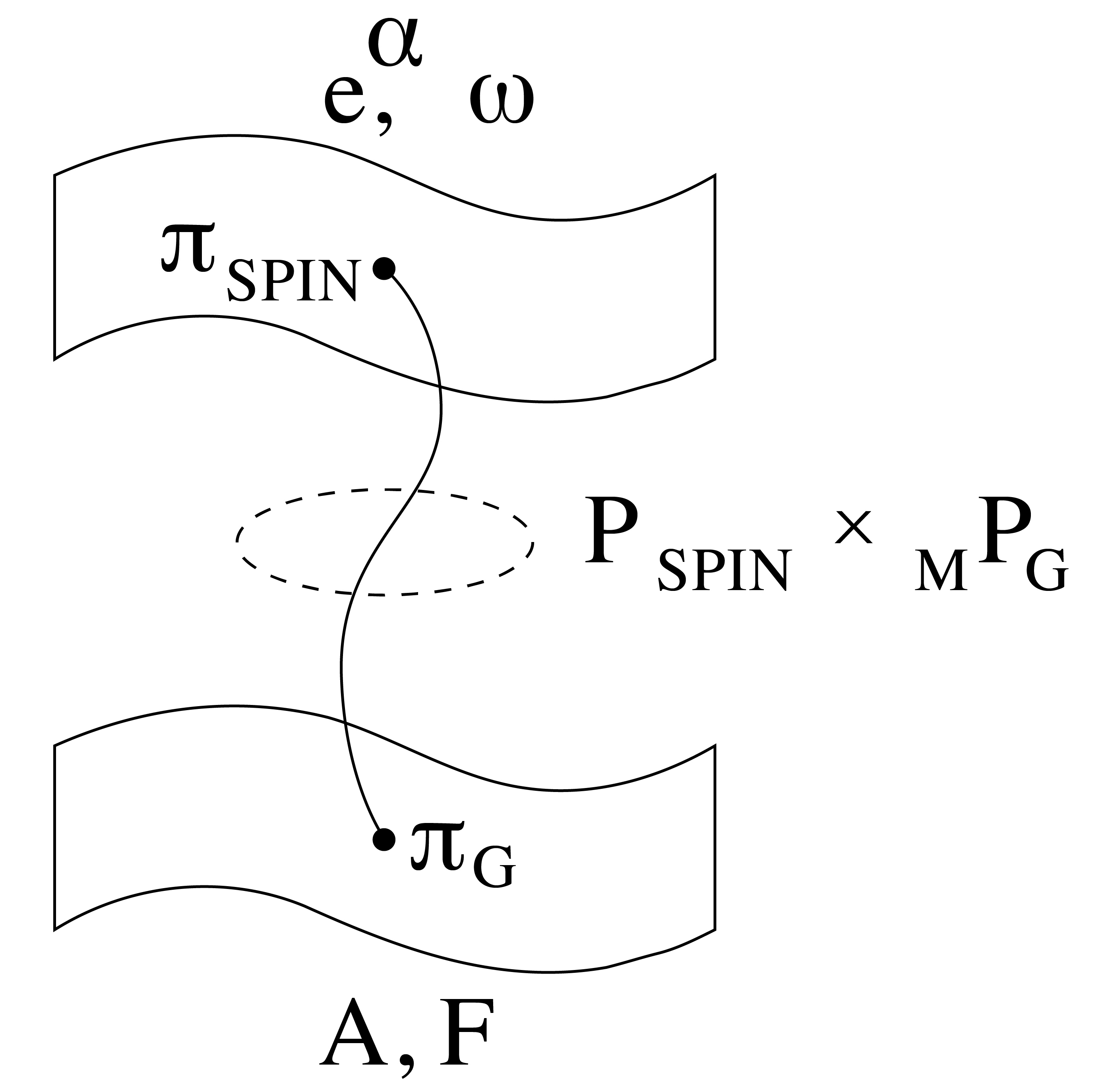}
    \caption{ The 
 top ribbon is the Spin frame bundle $\pi_{\mathrm{Spin}}:P_{\mathrm{Spin}}\to M$
  carrying the tetrad or coframe and spin connection $(e^\alpha,\omega)$;
  the bottom ribbon is the internal gauge bundle $\pi_G:P_G\to M$ carrying the gauge potential and curvature $(A,F)$.
  The dashed oval indicates the fiber product over a common base point $x\in M$.}
    \label{fig:prod-bundle}
\end{figure}

For the bundle geometry of the bundle $P_{Spin} \times _MP_G$ as connected by $\mathcal{F,A}$, we have $e^\alpha$ as the coframe one form that converts spacetime indices ($\mu,\nu,...$) to local Lorentz indices ($\alpha,\beta,...$). We have a set of one forms on $M$ as shown in Figure \ref{fig:prod-bundle}:
\begin{equation}
    e^\alpha=e_\mu^\alpha dx^\mu, \quad \alpha=0,1,2,3.
\end{equation}
The dual to the frame vector fields is $e_\alpha=e^\mu_\alpha\partial_\mu$ with
\begin{equation}
    e^\alpha(e_\beta)=\delta_\beta^\alpha, \quad e^\mu_\alpha e^\beta_\mu=\delta_\alpha^\beta, \quad e^\alpha_\mu e^\nu_\alpha=\delta_\mu^\nu.
\end{equation}
This builds the spacetime metric from the Minkowski metric $\eta_{\alpha\beta}=\text{diag}(-+++)$:
\begin{equation}
    h_{\mu\nu}=\eta_{\alpha\beta} e_\mu^\alpha e^\beta_\nu, \quad h^{\mu\nu}=\eta^{\alpha\beta} e^\mu_\alpha e_\beta^\nu, \quad \sqrt{|h|}=\text{det}(e^\alpha_\mu).
    \label{coframe1-forms}
\end{equation}
In this, $e^\alpha$ provides the soldering between the Spin bundle and the tangent bundle (TM) of the manifold $M$. 

For each point \(x\in M\), the~tangent space \(T_xM\) is the vector space of velocities of
smooth curves through \(x\) or equivalently, derivations on \(C^\infty(M)\) at \(x\).
The tangent bundle is the disjoint union of all tangent spaces, with~a smooth bundle structure:
\be
TM \;=\; \bigsqcup_{x\in M} T_xM,
\qquad
\pi_{TM}:TM\to M,\quad (x,v)\mapsto x.
\ee
If \(\dim M=n\), here \(n=4\), then each fiber \(T_xM\cong \mathbb{R}^n\).
In a chart \((x^\mu)\) on \(M\), a~tangent vector is \(v=v^\mu \partial_\mu\!\mid_x\),
and a point of \(TM\) has coordinates \((x^\mu,v^\mu)\). We have a vector field, which is a section of \(TM\):
\be
X:M\to TM, \qquad \pi_{TM}\circ X=\mathrm{id}_M,
\ee
with components \(X=X^\mu(x)\,\partial_\mu\).
The cotangent bundle \(T^*M\) has fibers of one-forms, so the~tetrad or coframe
\(e^\alpha=e^\alpha{}_\mu\,dx^\mu\) lives in \(\Omega^1(M)=\Gamma(T^*M)\).
The metric \(h\) is a section of \(S^2T^*M\) and identifies vectors or covectors through the musical isomorphisms.  To review, musical isomorphisms have the form:
\be
\sharp=\hat{g}:T^*M\to TM, \quad \flat=\hat{g}:TM\to T^*M,
\ee
these may also be referred to as mutually inverse isomorphisms, and~the (co)vectors obtained in this way are called metrically equivalent \cite{Music, Curvature}:
\be
X^\flat = h_{\mu\nu}X^\mu dx^\nu, \qquad
\alpha^\sharp = h^{\mu\nu}\alpha_\nu\,\partial_\mu .
\ee
We pick a local orthonormal basis \(e_\alpha=e_\alpha{}^\mu\partial_\mu\) of \(TM\).
The coframe \(e^\alpha=e^\alpha{}_\mu dx^\mu\) is its dual,~and
\be
h_{\mu\nu}=\eta_{\alpha \beta}\,e^\alpha{}_\mu\,e^\beta{}_\nu .
\ee
This is what we mean by the tetrad soldering \(P_{\mathrm{Spin}}\) to \(TM\);
pulling back the solder form on the spin bundle gives \(e^\alpha\), a~\(T^*M\)-valued~object. For \(\pi:M_{\mathbb C}\to M\) with \(z^\mu=x^\mu+i\,y^\mu\), the~tangent bundle of the complexification is \(TM_{\mathbb C}\).
An Ehresmann connection on \(\pi\) gives, near~\(y=0\), a~splitting:
\be
TM_{\mathbb C}\ \simeq\ \mathrm{Hor}\ \oplus\ \mathrm{Ver},\qquad
\mathrm{Ver}=\ker(d\pi),
\ee
so horizontal directions project to \(TM\) while vertical directions are along the
\(y\)-fibers.

If $s:M\to P_{Spin}$ is a local selection and $\vartheta^\alpha$ is the solder form, then $\vartheta^\alpha$ is the $\mathbb{R}^{1,3}$-valued one-form on the orthonormal frame or spin bundle that identifies tangent directions on $M$ with components in a chosen local Lorentz frame. It is the geometric device that connects the principal frame bundle to the base manifold's tangent~bundle.

We let $FM$ be the frame bundle of an $n$-manifold $M$, and we have a point $u\in FM$ which is a linear isomorphism $u:\mathbb{R}^n\to T_x M$, a~frame at $x=\pi(u)$. The~solder form $\vartheta\in\omega^1(FM);\mathbb{R}^n$ is defined by:
\begin{equation}
    \vartheta_u(X):=u^{-1}(d\pi_u(X))\in\mathbb{R}^n, \quad X\in T_uFM.
\end{equation}
So this means that we project a tangent vector $X$ at $u$ down to $T_xM$ through $d\pi$, then express it in the frame $u$ by applying $u^{-1}$. It has the property that if $X$ is tangent to the fiber, $d\pi(X)=0$, the~$\vartheta(X)=0$. For~the right action $R_g:FM\to FM, \quad g\in GL(n)$:
\begin{equation}
    (R_g)^*\vartheta=g^{-1}\vartheta.
\end{equation}
On the orthonormal frame bundle $P_{SO(1,3)},$ this holds with $g\in SO(1,3)$. For frames to the coframe on $e^\alpha$ on M, we chose a local section, $s : U \subset M \implies P_{SO(1,3)}$. We pull back the solder form:
\begin{equation}
    e^\alpha=s^*u^\alpha\in\Omega^1(U).
\end{equation}
These are the tetrad or coframe one-forms (\ref{coframe1-forms}). The relation to the spin bundle is shown when we let $\lambda:P_{Spin}\implies P_{SO(1,3)}$ be the 2-to-1 covering, where $\lambda$ is the bundle map that realizes the spin structure---the double cover from the Spin principal bundle to the orthonormal frame bundle. We pull $\vartheta$ back along $\lambda$ to get the spin bundle version:
\begin{equation}
\vartheta_{Spin}=\lambda^*_{Spin}\vartheta^\alpha.
\end{equation}
This transforms as a Lorentz vector under Spin(1,3) rotations:
\begin{equation}
    e^\alpha\to\Lambda_\beta^\alpha (x)e^\beta.
\end{equation}
Given the Spin connection $\omega_\beta^\alpha$ on $P_{Spin}$, the~Cartan structure equations are:
\begin{equation}
    T^\alpha=de^\alpha+\omega^\beta_\alpha\wedge e^\beta, \quad R_\beta^\alpha=d\omega_\gamma^\alpha+\omega_\gamma^\alpha \wedge \omega_\beta^\gamma,
\end{equation}
for Levi--Civita geometry $T^\alpha = 0$. For~completeness, we recall why the metric, Levi--Civita derivation used below is equivalent to the first-order tetrad or Palatini derivation when torsion vanishes. Given a coframe $e^a=e^a{}_\mu\,dx^\mu$ with inverse $e_a{}^\mu$, the Lorentzian metric is:
\begin{equation}
h_{\mu\nu}=\eta_{ab}\,e^a{}_\mu e^b{}_\nu,\qquad
\sqrt{|h|}=\det(e^a{}_\mu).
\end{equation}
Variations are related by:
\begin{equation}
\delta h_{\mu\nu}=\eta_{ab}\Bigl(e^a{}_\mu\,\delta e^b{}_\nu+e^a{}_\nu\,\delta e^b{}_\mu\Bigr).
\end{equation}
In the first-order Hilbert--Palatini formulation, the gravitational action can be written \mbox{as follows:}
\begin{equation}
S_{\rm HP}[e,\omega]=\frac{1}{4\kappa}\int_M \epsilon_{abcd}\,e^a\wedge e^b\wedge R^{cd}(\omega),
\end{equation}
where $R^{ab}(\omega)=d\omega^{ab}+\omega^a{}_c\wedge \omega^{cb}$ and $\epsilon_{abcd}$ is the Levi--Civita symbol in the local Lorentz frame. Varying independently in $(e^a,\omega^{ab})$ yields $\delta_\omega S_{\rm HP}=0\ \Rightarrow\ T^a:=de^a+\omega^a{}_b\wedge e^b=0$,
so the unique solution is the torsionless spin connection $\omega=\omega(e)$, inserting $\omega(e)$ into $\delta_e S_{\rm HP}=0$ gives the Einstein equation,
which is equivalent to the metric variation of the Einstein--Hilbert action $S_{\rm EH}[h]=\frac{1}{2\kappa}\int_M\sqrt{|h|}\,R(h)$.

In the torsionless sector that is relevant here, the~tetrad or Palatini and metric or Levi--Civita
formulations produce the same equations of motion and the same solution space modulo
local Lorentz gauge transformations of the~tetrad~\cite{DadhichPons2012}.

As a key summary, the~variational derivations for GR and Yang--Mills. We vary with respect to $h^{\mu\nu}$, the~contravariant metric, and~$A_\mu^A$. We set $F^A_{\mu\nu}=\partial_\mu A_\nu^A-\partial_\nu A_\mu^A+f^{A}{}_{BC}A_\mu^B A_\nu^C$,
$\langle F,F\rangle_h:=\kappa_{AB} F^A_{\mu\nu}F^{B\,\mu\nu}$, and~$D_\mu(\cdot)=\partial_\mu(\cdot)+[A_\mu,(\cdot)]$, we also assume either compact support of variations or the usual boundary~terms. We have the metric identities:
\begin{align}
\delta\sqrt{|h|} &= -\tfrac12\sqrt{|h|}\,h_{\mu\nu}\,\delta h^{\mu\nu},\\
\delta R &= R_{\mu\nu}\,\delta h^{\mu\nu}+\nabla_\mu\!\big(\nabla_\nu\,\delta h^{\mu\nu}-\nabla^{\mu}\delta h\big),
\quad \delta h:=h_{\alpha\beta}\delta h^{\alpha\beta}.
\end{align}
therefore:
\begin{equation}
\delta\!\big(\sqrt{|h|}R\big)=\sqrt{|h|}\,G_{\mu\nu}\,\delta h^{\mu\nu}
+\text{(total derivative)}\!,
\qquad G_{\mu\nu}:=R_{\mu\nu}-\tfrac12 h_{\mu\nu}R.
\end{equation}
For the gravitational part $S_{\rm EH}=\tfrac{1}{2\kappa}\int_M\!\sqrt{|h|}\,R$:
\begin{equation}
\delta S_{\rm EH}=\frac{1}{2\kappa}\int_M\!\sqrt{|h|}\,G_{\mu\nu}\,\delta h^{\mu\nu}
+\frac{1}{2\kappa}\int_{\partial M}\!\cdots
\end{equation}
The boundary term is cancelled by the Gibbons--Hawking--York term~\cite{York1972, GibbonsHawking1977}:
\be
S_{\rm GHY}=\frac{1}{\kappa}\int_{\partial M}\!\sqrt{|{\gamma}|}\,K,
\ee
for Dirichlet data on $h$. For~matter: 
\be
S_{\rm m}=\int_M\!\sqrt{|h|}\,\mathcal L_{\rm matter}
\ee 
we define:
\begin{equation}
T_{\mu\nu}:=-\frac{2}{\sqrt{|h|}}\frac{\delta S_{\rm m}}{\delta h^{\mu\nu}}.
\end{equation}
Stationarity $\delta S_{\rm EH}+\delta S_{\rm m}=0$ for arbitrary $\delta h^{\mu\nu}$ gives
\begin{equation}
 \; G_{\mu\nu}(h)=\kappa\,T_{\mu\nu}\;.
\end{equation}
We write the YM action in differential-form notation:
\begin{equation}
S_{\rm YM}=-\tfrac12\int_M\!\langle F\wedge{*}_h F\rangle
\;=\;-\tfrac14\int_M\!\sqrt{|h|}\,\kappa_{AB}\,F^A_{\mu\nu}F^{B\,\mu\nu} .
\end{equation}
Since $\delta F=D_A(\delta A)$, we have:
\begin{align}
\delta S_{\rm YM}
&=-\int_M\!\langle D_A\delta A\wedge{*}F\rangle
 \;=\;-\int_M\!d\langle \delta A\wedge{*}F\rangle+\int_M\!\langle \delta A\wedge D_A{*}F\rangle.
\end{align}
With $\delta A|_{\partial M}=0$ or by adding the natural boundary term, the~first term drops.
Matter coupling defines the gauge current by:
\begin{equation}
\delta S_{\rm m}\big|_{A}=\int_M\!\langle \delta A\wedge{*}J\rangle
\quad\Longleftrightarrow\quad
\frac{\delta S_{\rm m}}{\delta A_\mu^A}=\sqrt{|h|}\,J^{A\,\mu}.
\end{equation}
Stationarity $\delta(S_{\rm YM}+S_{\rm m})=0$ for arbitrary $\delta A$ yields the YM equations:
\begin{equation}
\; D_A({*}_h F)={*}_h J \;\qquad
\text{or in components }D_\mu F^{A\,\mu\nu}=J^{A\,\nu}\text{}.
\end{equation}
\textls[-15]{Varying $S_{\rm YM}$ with respect to $h^{\mu\nu}$ (using $\delta(\sqrt{|h|} F^2)
=\sqrt{|h|}\,[\, -\tfrac12 h_{\mu\nu}F^2 + 2\,F_{\mu\rho}F_\nu{}^{\rho}\,]\,\delta h^{\mu\nu}$) gives:}
\begin{equation}
T^{\rm YM}_{\mu\nu}
=\kappa_{AB}\!\left(F^A_{\mu\rho} F^{B}{}_{\nu}{}^{\rho}
-\tfrac14 h_{\mu\nu} F^A_{\rho\sigma}F^{B\,\rho\sigma}\right).
\end{equation}
Diffeomorphism invariance implies $ \nabla^\mu G_{\mu\nu}\equiv 0$, contracted Bianchi implies
\mbox{\(\nabla_\mu T^{\mu\nu}=0\)} on shell.
Gauge invariance implies $ D_A D_A({*}F)\equiv 0$ implies covariant current conservation
\(D_A J=0\).

In the Palatini or tetrad formalism~\cite{Palatini1919,HehlEtAl1976,Holst1996} one varies $(e^a,\omega^{ab})$ independently, and the~torsionless solution reproduces the Levi--Civita case. If~$M$ has a boundary, use
$S_{\rm GHY}$ and the YM boundary term $\int_{\partial M}\!\langle \delta A\wedge{*}F\rangle$ to enforce Dirichlet~data.

\section{Real Slice~Geometry}

We will now explain real-slice geometry as a finer bundle to~show how the real slice geometry encodes symmetries in HUFT. By~a finer bundle over \(M\) we mean an enlargement of a given bundle
\(q:P \to  M\) by adjoining an extra fiber \(E_y \to  M\), encoding the
imaginary directions \(y\) from the complexification \(M_{\mathbb{C}}\):
\be
\widehat P \;:=\; P \times_M E_y \;\longrightarrow\; M,
\qquad
\pi_1:\widehat P\to P \text{ over } \mathrm{id}_M,
\ee
so that fiberwise \(\widehat P_x \cong P_x \times (E_y)_x\) and \(\pi_1\) simply
forgets \(y\). In~our setting \(P=P_{\mathrm{tot}}=P_{\mathrm{Spin}}\times_M P_G\),
so \(\widehat P=(P_{\mathrm{Spin}}\times_M P_G)\times_M E_y\).
This is not a reduction in the structure group nor a refinement of an atlas, but~it is an enrichment of the fiber used to keep track of the \(y\)-directions while all
fields and dynamics are evaluated on the real slice $M$. We let $M$ be a smooth oriented, time-orientable spin $4$-manifold, the~real slice. We have $M_{\mathbb{C}}$ as a complexification with coordinates $z^\mu=x^\mu+i y^\mu$ and
projection $\pi:M_{\mathbb{C}}\to M$, $\pi(x,y)=x$. The~fibers $Y_x:=\pi^{-1}(x)$ are real $4$-planes isomorphic to $\mathbb R^4$, defining a rank-$4$ real vector bundle $E_y\to M$ with total space canonically identified near $y=0$ with a neighborhood of $M\subset M_{\mathbb{C}}$.

We have $P_{Spin}\to M$ as the $Spin(1,3)$ frame bundle and $P_G\to M$ a principal
$G$-bundle with $G= SU(3)\times SU(2)\times U(1)$. Now, we define the product bundle
$P_{tot}:=P_{Spin}\times_M P_G$ with structure group $H:=Spin(1,3)\times G$.
We refine this by adjoining the $y$-fiber:
\be
\widehat{P}\;:=\;P_{tot}\times_M E_y \;\;\longrightarrow\;\; M,
\ee
whose points encode local Lorentz frames, internal gauge frames, and~the imaginary directions $y$ of the complexified manifold. Now, we choose an $H$-connection $\mathcal{A}=(\omega,A)$ on $P_{\mathrm{tot}}$ and an Ehresmann connection
on $\pi:M_{\mathbb{C}}\to M$ that splits $T M_{\mathbb{C}}\simeq \mathsf{Hor}\oplus \mathsf{Ver}$ near $y=0$.
Now, we write the unified curvature:
\be
\mathcal{F}\;=\;\mathrm{d}\mathcal{A}+\mathcal{A}\wedge\mathcal{A}
\;=\; R\oplus F,\qquad
R\in\Omega^2(M,\mathfrak{spin}(1,3)),\;\;F\in\Omega^2(M,\mathrm{ad}P_G),
\ee
and let $h\in\Gamma(\text{Sym}^2 T^*M)$ be a Lorentzian metric, used for index operations and Hodge duals.
On $M_\mathbb{C}$ we consider a Hermitian field:
\be
g \;=\; h \;+\; i\,B,\qquad B\in\Omega^2 \big(M,\mathrm{ad}P_G\big),
\label{eq:hermition}
\ee
and identify on the real slice $B=F$ either by definition or via a holomorphic
compatibility term enforcing $B-F=0$ in the equations of motion.
The base metric $h$ transforms under
$\text{Diff}(M)$ as its Levi--Civita and spin connection $\omega$ is the $Spin(1,3)$ part of $A$. The~curvature $R$ is the gravitational field strength; all contractions, Hodge duals,
and stress-energy are built using $h$. The~internal symmetry $G=SU(3)\times SU(2)\times U(1)$
acts on $P_G$; the unified gauge potential $A=A^{(3)}\oplus A^{(2)}\oplus A^{(1)}$ and
field strength $F=F^{(3)}\oplus F^{(2)}\oplus F^{(1)}$ live in
$\Omega^2(M,\mathrm{ad}P_G)$. On~the real slice, the~antisymmetric part of the Hermitian
metric equals this curvature: $g_{[\mu\nu]}\equiv B_{\mu\nu}=F_{\mu\nu}$. For the Finer bundle with $y$ inside, the $y$–fiber $E_y$ records the imaginary
directions of $M_\mathbb{C}$. The~Ehresmann splitting ties variations in $y$ to horizontal transport
on $M$. In~the holomorphic description, the~mixed $(x,y)$-geometry packages the gauge sector
into the antisymmetric two-form $B$ while preserving standard spacetime geometry in $h$. By~Noether I and II, we obtain the single invariant~action.
A single $\text{Diff}(M)\times G$-invariant~action:
\be
S[h,\mathcal{A},\Psi]
=\int_M \sqrt{|h|}\left(\frac{1}{2\kappa}R(h)-\frac14\langle F,F\rangle_h
+ \mathcal{L}_{\rm matter}(\Psi;h,\mathcal{A})\right),
\ee
yields, by~variation:
\be
G_{\mu\nu}(h)=\kappa\,T_{\mu\nu}(F,\Psi;h),\qquad D_A(*_hF)=J_{\rm matter},
\ee
and by Noether’s second theorem, the unified Bianchi identities $D_\omega R=0$, $D_A F=0$, together with covariant current conservation $\nabla_\mu T^{\mu\nu}=0$, $D_A J=0$. The~same symmetry explains the
conservation laws. Writing $G=SU(3)\times SU(2)\times U(1)$ with Lie algebra
$\mathfrak g=\mathfrak{su}(3)\oplus\mathfrak{su}(2)\oplus\mathfrak{u}(1)$ and Killing form
$\kappa$, we decompose
\be
A = A_\mu dx^\mu = A^{(3)a}_\mu T^{(3)}_a\,dx^\mu
\oplus A^{(2)i}_\mu T^{(2)}_i\,dx^\mu
\oplus A^{(1)}_\mu Y\,dx^\mu,
\ee
\be
F = dA + A\wedge A = F^{(3)}\oplus F^{(2)}\oplus F^{(1)},\qquad
\langle F,F\rangle_h = \kappa_{AB} F^A_{\mu\nu}F^{B\,\mu\nu}.
\ee
On the real slice $y=0$:
\be
\;g_{[\mu\nu]} \;\equiv\; B_{\mu\nu} \;=\; F_{\mu\nu}^{(3)}\oplus F_{\mu\nu}^{(2)}\oplus F_{\mu\nu}^{(1)}\;,
\ee
and all index operations use only $h^{\mu\nu}$. Thus, the~antisymmetric sector encodes the entire SM gauge curvature, while the symmetric sector encodes spacetime~geometry.

The refined bundle $\widehat{P}$ packages diffeomorphisms through $h,\omega$ and SM gauge
symmetries via $A,F$ and therefore $g_{[\mu\nu]}$, all governed by one symmetry group
$\text{Diff}(M)\times G$ and one invariant action. This realizes Einstein’s symmetry-first mandate and Noether’s criterion that conservation and identities follow from the same~symmetry.

\begin{figure}[H]
    
    \includegraphics[width=0.4\linewidth]{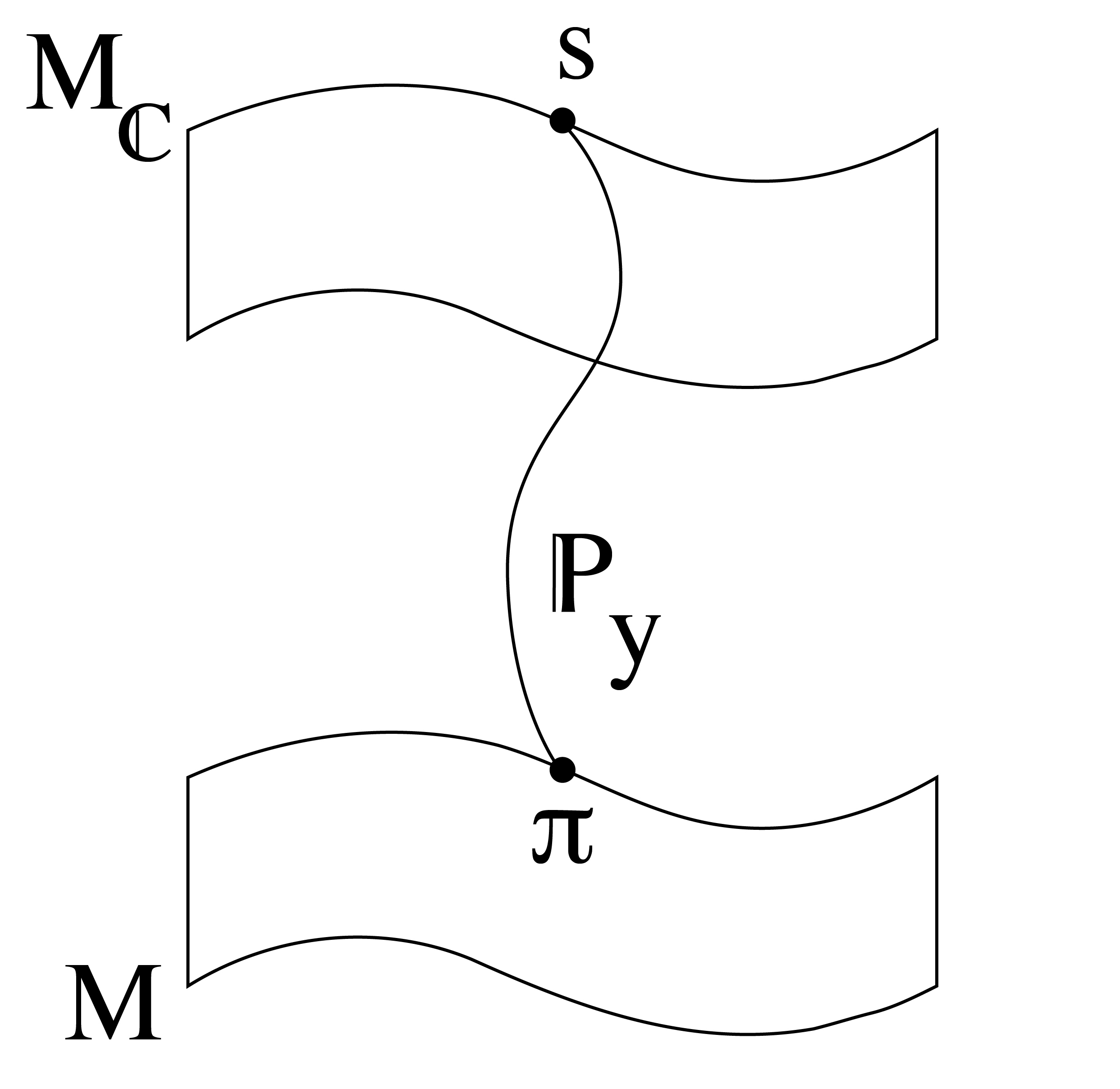}
    \caption{The upper
 ribbon is $M_{\mathbb C}$ with coordinates $z^\mu=x^\mu+i\,y^\mu$,
   the lower ribbon is the real slice $M$.
  The projection $\pi:M_{\mathbb C}\!\to M$ sends $(x,y)\mapsto x$.
  The black dots mark the point $s_0(x)=(x,0)\in M_{\mathbb C}$ (label $s$) and its projection $x\in M$ (label $\pi$), so that $\pi\circ s_0=\mathrm{id}_M$.
  The label $\mathbb{P}_y$ denotes the rank-$4$ real vector bundle $E_y\to M$ of imaginary directions:
  for each $x\in M$, the~fiber $Y_x=\pi^{-1}(x)\cong\mathbb R^4$ carries coordinates $y^\mu$.}
    \label{fig:real-slice}
\end{figure}

The diagram Figure \ref{fig:real-slice} shows the complexification $\pi:M_{\mathbb C}\!\to M$, with~the upper ribbon $M_{\mathbb C}$  and the lower ribbon is the real slice $M$. The~point labelled $s$ is the image of the zero section $s_0:M\hookrightarrow M_{\mathbb C}$, $s_0(x)=(x,0)$; the lower point labelled $\pi$ is the basepoint $x\in M$, so $\pi\!\circ s_0=\mathrm{id}_M$. The~curved line between them indicates the fiber $Y_x:=\pi^{-1}(x)$ over $x$, which is a real $4$-plane isomorphic to $\mathbb R^4$ and coordinatized by the imaginary directions $y^\mu$. We denote by $\mathbb{P}_y$ in the figure our $E_y$ in the text and the rank–$4$ real vector bundle $E_y\to M$ whose fiber at $x$ is $Y_x$; near $y=0$ its total space is canonically identified with a neighborhood of $M\subset M_{\mathbb C}$. This bookkeeping of the $y$-directions lets us speak of a refined finer bundle over $M$ that remembers the imaginary displacements while all fields are ultimately evaluated on the real~slice.

\section{Symmetry Completeness and Its Physical~Equivalence}

As a sanity check, we will show that we recover symmetry completeness and explore its physical equivalence.
We have $M$ as an oriented, time-orientable spin $4$–manifold and let $P_{\text{Spin}} \to  M$ be the $\text{Spin}(1,3)$ frame bundle and
$P_G \to  M$ a principal $G$-bundle with $G=SU(3)\times SU(2)\times U(1)$.
Set $P_{\text{tot}}=P_{\text{Spin}}\times_M P_G$ with structure group $H=\text{Spin}(1,3)\times G$.
Now, let $h\in\Gamma(\text{Sym}^2T^*M)$ be a Lorentzian metric and
$\mathcal{A}=(\omega,A)$ an $H$-connection with curvature
$\mathcal{F}=\mathrm{d}\mathcal{A}+\mathcal{A}\wedge\mathcal{A}=(R,F)$, where $R$ is the Riemann curvature $2$-form
and $F$ the Yang--Mills field strength.
On the complexified bundle of $(0,2)$-tensors, define the Hermitian field as in Equation (\ref{eq:hermition})
and on the real slice impose the compatibility $B=F$ either as a definition
or via a holomorphic penalty term enforcing $B-F=0$ in the equations of~motion. We consider the single $\text{Diff}(M)\times G$-invariant action:
\begin{equation}\label{eq:UnifiedAction}
S[h,\mathcal{A},\Psi]
=\int_M \sqrt{|h|}\,\Big(\tfrac{1}{2\kappa}R(h)
-\tfrac{1}{4}\,\langle F,F\rangle_h
+ \mathcal{L}_{\rm matter}(\Psi;h,\mathcal{A})\Big).
\end{equation}
where $\langle\cdot,\cdot\rangle_h$ is the fiberwise inner product induced by $h$
and the Killing form on $\mathfrak g$, and~$\Psi$ denotes matter fields. Here, $\mathcal{A}=(\omega,A)$ with curvature $\mathcal{F}= \mathrm{d}\mathcal{A}+\mathcal{A}\wedge\mathcal{A}=(R,F)$, so $D_{\mathcal{A}}=(D_\omega,D_A)$ on $\mathfrak{spin}(1,3)\oplus\mathfrak{g}$.

\begin{Theorem}\label{thm:equiv}
For the data above, the~following hold on the real slice $B=F$:
\begin{enumerate}
\item[(i)] $S$ is invariant under $\text{Diff}(M)\times H$.
\item[(ii)] Noether II for $\text{Diff}(M)\times G$ yields the unified Bianchi identities
$D_\omega R=0$ and $D_A F=0$; Noether I yields the covariant conservations
$\nabla_\mu T^{\mu\nu}=0$ and $D_A J=0$.
\item[(iii)] The Euler--Lagrange equations of $S$ are
\begin{equation}\label{eq:EOM}
G_{\mu\nu}(h)=\kappa \big(T^{\rm YM}_{\mu\nu}+T^{\rm matter}_{\mu\nu}\big),
\qquad
D_A(*_h F)=J_{\rm matter},
\end{equation}
with the standard Yang–Mills stress tensor
$T^{\rm YM}_{\mu\nu}= {\rm tr} \left(F_{\mu\alpha}F_\nu{}^{\alpha}
-\tfrac14 h_{\mu\nu}F_{\alpha\beta}F^{\alpha\beta}\right)$
and gauge current $J$ defined from $\mathcal{L}_{\rm matter}$ by minimal coupling.
\item[(iv)] (Classical equivalence of physics) The map
\be
\Phi:\ (h,A,\Psi)\ \longmapsto\ (g=h+iF,\ \mathcal{A}=(\omega(h),A),\ \Psi),
\ee
induces a bijection between solution spaces modulo $\text{Diff}(M)\times G$:
\be
\mathfrak{Sol}_{\rm EYM+matter}/(\text{Diff}\times G)
\;\cong\;
\mathfrak{Sol}_{\rm unified\ real\ slice}/(\text{Diff}\times G).
\ee
Hence, all classical observables and their conservation laws coincide.
\item[(v)] (Quantum equivalence, formal) If one includes a gauge-invariant penalty
$\frac{\lambda}{2} \int \sqrt{|h|}\,\langle B-F,B-F\rangle_h$ and integrates out $B$,
then, for~$\lambda \to \infty$, the~generating functional reduces to that of the standard
Einstein--Yang--Mills and matter theory, so perturbative
correlators and $S$–matrix elements agree.
\end{enumerate}
\end{Theorem}

Under $H$-gauge transformations, $F\mapsto g^{-1}Fg$ and the Killing form is
$\text{Ad}$-invariant, so $\mathrm{tr}(F\wedge *_h F)$ is gauge-invariant. Each term in \eqref{eq:UnifiedAction}
is a scalar density; hence, it is $\text{Diff}(M)$-invariant. For~a connection $A$ on a principal $H$-bundle,
$D_{A}F=0$ is the unified Bianchi identity, which splits as $D_\omega R=0$ and
$D_A F=0$. Diffeomorphism invariance implies the contracted Bianchi identity
$\nabla_\mu G^{\mu\nu}=0$ and hence $\nabla_\mu T^{\mu\nu}=0$ on shell $G$-invariance implies $D_A J=0$. Varying $A$ with $h$ fixed gives:
\be
\delta S_{\rm YM} = -\tfrac12\int \sqrt{|h|}\,\langle D_A(*_h F),\delta A\rangle,\qquad
\delta S_{\rm matter}=\tfrac12\int \sqrt{|h|}\,\langle J_{\rm matter},\delta A\rangle,
\ee
so $D_A(*_h F)=J_{\rm matter}$.
Varying $h$ yields:
\be
\delta S = \tfrac12\int \sqrt{|h|}\,\big(G_{\mu\nu}-\kappa\,T_{\mu\nu}\big)\,\delta h^{\mu\nu},
\ee
with $T_{\mu\nu}=T^{\rm YM}_{\mu\nu}+T^{\rm matter}_{\mu\nu}$, giving the Einstein~equation.

Given any EYM\(+\)matter solution $(h,A,\Psi)$,
define $(g,A,\Psi)$ by $g=h+iF$ and \linebreak  $\mathcal{A}=(\omega(h),A)$. Then, $(g,A,\Psi)$ solves the unified Euler--Lagrange system on the real slice because \eqref{eq:EOM} hold, and~$B=F$ by construction. Any unified real-slice solution $(g,A,\Psi)$ has $g=h+iB$ with $B=F$ from the $B$–equation, and~its $(h,A,\Psi)$ satisfies \eqref{eq:EOM}.
Both constructions are natural with respect to $\text{Diff}(M)\times G$, so they descend to a bijection of quotient solution spaces. Classical observables built from $(h,A,\Psi)$, such as fluxes, charges, and stress-energy, coincide with those built from $(g,A,\Psi)$ restricted to the real~slice. \mbox{We add:}
\be
S_\lambda[B,A;h]=\tfrac{\lambda}{2}\int \sqrt{|h|}\,\langle B-F,B-F\rangle_h,
\ee
the $B$-integral is Gaussian:
\be
\int   D B \, e^{\,i S_\lambda[B,A;h]}\;\propto\; (\det \lambda)^{-1/2}\,
e^{\,i\cdot 0}\,\delta[B-F]\ \xrightarrow{\ \lambda\to\infty\ }\ \delta[B-F].
\ee
Thus, the~unified partition function reduces to the standard EYM\(+\)matter one up to an overall constant, with~identical gauge-fixing or ghost structure. Gauge and diffeo invariance and BRST remain as in the standard~theory.

HUFT is a unification of geometric data and variational origin rather than a claim of a
new nonperturbative quantum-gravity principle. Classically, the~real-slice theory is equivalent to Einstein--Yang--Mills and matter
(Theorem~1), so HUFT makes the same classical gravitational predictions as GR coupled
to the same gauge/matter content. Perturbatively, the~formal quantum equivalence statement means that when the
compatibility field $B$ is treated as auxiliary and integrated out, the~generating functional
reduces to that of standard EYM+matter up to an overall constant and identical gauge-fixing, so the perturbative spectrum and correlators are unchanged. For~the UV completion and quantum-gravity aspect, if~one seeks an ultraviolet-softened, BRST-compatible completion, the~only allowed modifications consistent with the assumed
local symmetries are covariant entire-function form factors $F(D^2/M_\ast^2)$ in the kinetic
operators as encoded in Theorem~2. In~that sense, HUFT provides a symmetry-organized arena for discussing UV-finite nonlocal extensions of gravity+gauge theory, while remaining IR-equivalent to the local~theory.

The consequences of this are that we have the same symmetries; $\text{Diff}(M)\times G$ invariance, Bianchi identities, and~covariant current and stress-energy conservation are identical to GR\(+\)YM. The~same observables, such as charges, fluxes, and~classical predictions, match, so perturbative quantum correlators agree when $B=F$ is enforced. The~unified packaging, gravity $R$ and gauge $F$, are components of one curvature $\mathcal{F}$ of one connection $\mathcal{A}$ on one bundle $P_{\mathrm{tot}}$.

\begin{Theorem}
Let $(M,h)$ be a real Lorentzian $4$-manifold admitting a complexification $M_\mathbb{C}$. Let $P_{\mathrm{tot}}\to M$ be a principal $H$-bundle with $H=\mathrm{Spin}(1,3)\times G$ where $G$ is compact, reductive, and~contains the SM gauge group. Consider  fields $\big(h_{\mu\nu}, \mathcal{A}=(\omega,A_{\mathrm{YM}}), \Psi, \Phi, g_{\mu\nu}\big)$ with
\be
g_{\mu\nu}=h_{\mu\nu}+i\,B_{\mu\nu}\quad\text{(Hermitian packaging on $M_\mathbb{C}$)}.
\ee
Assume the dynamics are invariant under the automorphism group
\be
\mathrm{Aut}(P_{\mathrm{tot}})=\mathrm{Diff}(M)\ltimes \mathrm{Gau}(P_{\mathrm{Spin}} \times_M P_G),
\ee
implemented off–shell by a nilpotent BRST differential $s$ (Noether II setting). The~Lagrangian density is a diffeo and gauge-natural polynomial in the fields and their covariant derivatives, and~may depend on the covariant d’Alembertian $D^2$ only through an entire functional calculus $F(D^2/M_*^2)$ that commutes with $s$ and all covariant derivatives. In~the limit $M_* \to \infty$, the~quadratic operators for $(h,A,\Psi,\Phi)$ reduce to second order in derivatives, with~indices raised and lowered solely by $h$. There is reflection positivity on the Euclidean slice as there are no extra propagating poles other than those of $(h,A,\Psi,\Phi)$. There are no additional massless higher-spin-gauge fields beyond spin–2 $h$ and spin–1 connections; these interactions are marginal or relevant in the IR. The~local BRST anomaly cohomology $H^{1,4}(s\mid d)$ vanishes for the chosen matter representation, with no gauge or diffeo~anomalies.

Then, modulo $s$-exact terms, total derivatives, and~higher-dimension operators are suppressed by $M_*$, and the~most general action is, after~field redefinitions, equivalent to
\be
S= \int_M \sqrt{|h|}\Big[\tfrac{1}{2\kappa}R(h)-\tfrac14\langle F_{\mathrm{YM}},F_{\mathrm{YM}}\rangle_h
+\overline{\Psi}(i\slashed{D}-Y[\Phi])\Psi+\tfrac12(D\Phi)^2 - V(\Phi)\Big]
+ S_{\mathrm{top}},
\ee
with the following consequences; BRST covariance implies the antisymmetric piece $B_{\mu\nu}$ transforms as a curvature two–form; hence, on the real slice,
\be
B \equiv F_{YM}, \,\,\,\text{as elements of} \,\,\,\,\ \Omega^2(M,\mathrm{ad}\,P_G),
\ee
where scalar contractions use $\kappa(\cdot,\cdot)$, so $g_{\mu\nu}=h_{\mu\nu}+iB_{\mu\nu}$ is a kinematic packaging and does not introduce an independent two-form gauge sector. Preserving the full local symmetry off–shell to all loops restricts UV softening to entire functions $F(D^2/M_*^2)$ at covariant kinetic operators and consistently at vertices. This renders perturbation theory UV–finite while the IR ($M_* \to \infty$) recovers local EYM$+$matter. Noether II yields the unified Bianchi and Slavnov--Taylor \mbox{identities, and}
\be
D_\mu J^\mu_a=0,\qquad \nabla_\mu T^{\mu\nu} = F^{\nu\rho}_a\,J^a_\rho,
\ee
with $T^{\mu\nu}$ the Belinfante tensor built from $h$. Up~to $S_{\mathrm{top}}$, Euler, Pontryagin, $\theta$–terms, no further diffeo or gauge-natural, unitary, second-order couplings exist.
\end{Theorem}

In this section, we specify the geometric data $P_{\mathrm{tot}}=P_{\mathrm{Spin}}\times_M P_G$ with structure group $H=\mathrm{Spin}(1,3)\times G$, a~single $H$-connection $\mathcal{A}=(\omega,A)$ with unified curvature $\mathcal F=(R,F)$, and~a Lorentzian metric $h$. The~master action Equation~\eqref{eq:master-action} is written and shown to be $\mathrm{Diff}(M)\times H$-invariant by construction. From~the single bundle identity $D_{\mathcal{A}}\mathcal{F}=0$, we then derive the split into the Riemann and Yang--Mills Bianchi identities, $D_\omega R=0$ and $D_A F=0$, thereby establishing the unified Noether–II content. Variation with respect to $A$ yields the Yang--Mills equations $D_A(*_h F)=J_{\rm matter}$ Equation~\eqref{eq:YM}, while variation with respect to $h$ yields Einstein’s equations sourced by the Yang--Mills and matter stress tensors Equation~\eqref{eq:Einstein}. Therefore, a single invariant action produces both sectors’ dynamics and their identities, exactly in Noether’s~sense.

Within these principles, there are no additional diffeomorphism- and gauge-natural, unitary second-order couplings in four dimensions beyond Einstein--Hilbert, Yang--Mills, and~minimal matter terms, as anything else is either purely topological, Euler, Pontryagin, $\theta$-terms or higher-dimension and therefore suppressed. Attempts to add an independent antisymmetric metric mode, extra massless higher-spin fields, or~non-natural derivative couplings either break the symmetry structure, introduce ghosts or extra poles, or~run afoul of anomaly constraints. Thus, in~this construction, SM $+$ GR is not an arbitrary choice but the unique possibility consistent with the Einstein--Noether program, and~the Hermitian geometry makes that uniqueness explicit by identifying the imaginary or antisymmetric sector
of the Hermitian packaging tensor with an adjoint-valued two-form and enforcing real-slice compatibility:
\begin{equation}
\mathrm{Im} (g_{\mu\nu})=\ell_*^{2}\,B_{\mu\nu},\qquad B\in\Omega^{2}(M,\mathrm{ad}\,P_G),
\end{equation}
\textls[-15]{The compatibility mechanism either imposed kinematically or derived dynamically enforces}:
\begin{equation}
B=F \quad \text{on the real slice } \, M,
\end{equation}
so that the Hermitian field $g=h+i\,\ell_*^{2}B$ is a packaging of $(h,F)$ rather than an additional
propagating tensor sector.
In particular, one should not write $g_{[\mu\nu]}=F_{\mu\nu}$ without specifying the projection
to the imaginary/antisymmetric part and the scale $\ell_*$.

\section{Field Geometry and Global Spatial~Topology}
\label{sec:shape-topology}

In HUFT, all fields live on the product principal bundle:
\be
P_{\mathrm{tot}} \;=\; P_{\mathrm{Spin}}\times_{M} P_G
,\qquad\text{with structure group},\qquad
H=\mathrm{Spin}(1,3)\times G,
\ee
with a single $H$-connection $\mathcal{A}=(\omega,A)$ whose curvature
$\mathcal{F}=d\mathcal{A}+\mathcal{A}\!\wedge\!\mathcal{A}=(R,F)$ splits into spacetime curvature $R$ and Yang--Mills curvature $F$.
On the complexified tangent bundle, we package the metric as a Hermitian field $g=h+iB$.
On the real slice $y=0$ used for physics, the~antisymmetric piece $B$ is not independent, as
it is identified with the gauge curvature two-form, $B\equiv F$, so the gauge-field shape is literally the two-form curvature living in the internal~fiber.

The internal fiber and vacuum manifold comes from us adopting the minimal simple gauge group:
\be
G=\mathrm{SU}(5),\qquad \dim G =24,
\ee
whose maximal torus or Cartan is $T^4=\mathrm{U}(1)^4$. The~electric and hypercharge assignments live along these four commuting directions, the~weights of the~Cartan. Spontaneous symmetry breaking proceeds in two steps:
\begin{align*}
\text{(GUT)}\qquad & \mathrm{SU}(5)\;\longrightarrow\;
\mathrm{SU}(3)_c\times \mathrm{SU}(2)_L\times \mathrm{U}(1)_Y,\\[2pt]
\text{(EW)}\qquad & \mathrm{SU}(2)_L\times \mathrm{U}(1)_Y
\;\longrightarrow\; \mathrm{U}(1)_{\mathrm{EM}}.
\end{align*}
The GUT vacuum manifold is the flag space:
\be
\mathcal{M}_{\mathrm{GUT}}
\;=\;
\frac{\mathrm{SU}(5)}{\mathrm{SU}(3)\times \mathrm{SU}(2)\times \mathrm{U}(1)},
\qquad \dim \mathcal{M}_{\mathrm{GUT}}=12,
\ee
and the electroweak vacuum manifold is:
\be
\mathcal{M}_{\mathrm{EW}}
\;\simeq\;
\frac{\mathrm{SU}(2)\times \mathrm{U}(1)}{\mathrm{U}(1)_{\mathrm{EM}}}
\;\cong\; S^3.
\ee
These spaces encode which gauge directions are unbroken versus broken, massless vs, massive,~respectively. Throughout this paper, we take spatial slices to be:
\be
\Sigma_3 \;\cong\; \mathbb{R}^3,
\ee
such as spatially flat and simply connected, in line with current cosmological evidence for near-flat geometry. This choice fixes boundary conditions and mode expansions but does not affect the local variational derivation of the field equations or the unification~mechanism.

Other constant-curvature topologies remain compatible with the framework, such as a three-torus $T^3$, where periodic identifications on three cycles lead to discrete comoving momenta and possible global Aharonov--Bohm phases for gauge holonomies around non-contractible loops. Defect classification uses the same homotopy groups, but~spatial infinity is replaced by large embedded two-tori, two-spheres, or~by cycle representatives. A~three-sphere $S^3$, finite spatial volume, harmonic analysis uses $S^3$ eigenmodes. Large-radius $S^2\subset S^3$ still serves to define monopole charges; instanton number is unchanged since it lives in $\pi_3(G)$. But~in all cases, the~internal geometry, fiber $G$, Cartan $T^4$, vacuum manifolds, and~the identification $B\equiv F$ on the real slice are unchanged, as~only global boundary conditions and spectral discreteness~differ.

Gauge- and BRST-covariant entire-function regulators $F(D^2/M_*^2)$ act as an ultraviolet texture filter, exponentially damping modes with $|p|\gg M_*$ without introducing new poles. Operationally, correlators are band-limited on scales shorter than $M_*^{-1}$, so fine-grained field wrinkles are smoothed while locality, gauge identities, and~unitarity are~preserved.

So, in HUFT, the universe’s field shape in this framework is a single connection on $P_{\mathrm{Spin}}\times_M P_G$ whose curvature splits into gravity and gauge parts, an~internal $\mathrm{SU}(5)$ fiber containing a Cartan $T^4$ that organizes charges, vacuum manifolds $\mathcal{M}_{\mathrm{GUT}}$; flag space and $\mathcal{M}_{\mathrm{EW}}\cong S^3$ that determine massless vs.\ massive directions, topological sectors classified by maps from $S^2$ and $S^3$ into these spaces,
and UV-smooth field textures controlled by $M_*$. For~definiteness, we fix $\Sigma_3\simeq\mathbb{R}^3$,
while noting that $T^3$ or $S^3$ spatial topologies are also compatible and leave the unification structure~intact.

In the present formulation, HUFT requires that the real slice \(M\) be an oriented, time-orientable spin four-manifold supporting the product principal bundle
\(P_{\mathrm{tot}}=P_{\mathrm{Spin}}\times_M P_G\) with a single \(H=\mathrm{Spin}(1,3)\times G\) connection \(\mathcal A=(\omega,A)\)
and curvature \(\mathcal F=R\oplus F\). On~the real slice, the~antisymmetric metric piece satisfies \(B\equiv F\), so the gauge sector is encoded directly in the fiber curvature.
Chern--Weil matching and anomaly cancellation fix the normalization of \(\langle\!\langle F\wedge F\rangle\!\rangle\)
and select the minimal internal group \(G=\mathrm{SU}(5)\).
These are global bundle constraints on \(M\), but~they do not determine the global topology of spacelike slices \(\Sigma_t\).

Accordingly, we do not commit the theory to a unique \(\Sigma_t\), but common choices such as
\(\mathbb{R}^3\), \(T^3\), or~\(S^3\) are all compatible, provided they admit the required spin structure and principal \(G\)-bundle. Physical consequences of different \(\Sigma_t\) arise via boundary conditions
such as discrete momenta on \(T^3\), finite-volume harmonics on \(S^3\), and~possible global holonomies, not from the unification mechanism itself. Any further selection among allowed topologies would have to
come from cosmological initial data or an extension of the action that weights topology~classes.

\section{Quantum Mechanics and Quantum Field Theory from~HUFT}

Now, to~show compatibility with quantum mechanics, we will show that the Born rule, Veltman condition, and~Dirac equation come out naturally from fibre bundles in HUFT~\cite{Gleason1957,BuschLahtiBook,ZurekEnvariance, Veltman1981}. First, recall that $(M,g_{\mu\nu})$ is the complex Hermitian spacetime of HUFT with real slice $M\subset M_\mathbb{C}$.
Matter fields live in a complex Hermitian vector bundle:
\be
\pi:E\to M,\qquad \text{rank}(E)=d\in\mathbb{N},
\ee
associated with a principal bundle with structure group $G\times U(1)$.
On the real slice, there is a positive Hermitian fiber metric:
\be
h_x: E_x\times E_x\to\mathbb{C},\quad h_x(\cdot,\cdot)=\langle\cdot,\cdot\rangle_x,
\ee
and a unitary connection $D=\mathrm{d}+{\cal A}$ compatible with $h$.
A pure quantum state at $x\in M$ is a ray \([\psi_x]\in\mathbb{P}(E_x)\), such as a nonzero vector modulo the $U(1)$ fiber phase. A~ray $[\psi]\in\mathbb P(E_x)$ is the $1$-dimensional subspace $\{\lambda\psi:\lambda\in\mathbb C^\times\}$; physical predictions depend only on $[\psi]$, equivalently on $\Pi_{[\psi]}=|\psi\rangle\langle\psi|/\langle\psi,\psi\rangle$. Observable $A$ is a Hermitian bundle endomorphism $A\in\Gamma(\mathrm{End}(E))$; at $x$, it has the spectral resolution
\(
A_x=\sum_i a_i\,\Pi_{i,x}
\)
with orthogonal projectors $\Pi_{i,x}\in \mathrm{End}(E_x)$.

Our goal is to construct a probability assignment with
\(
\mu_x:\{\text{projectors on }E_x\}\to[0,1]
\)
meaning that we assign a probability number to every projector or outcome at $x$, such that for the outcome subspace $\mathrm{Im}\,\Pi_{i,x}$, we have
\be
\mu_x(\Pi_{i,x}\,|\, [\psi_x]) \;=\; \|\Pi_{i,x}\psi_x\|_{h_x}^2/\|\psi_x\|_{h_x}^2.
\ee

We fix $x\in M$ and suppress the subscript $x$.
Let $\mathcal{P}$ be the lattice of orthogonal projectors on $E_x$. We assume $\mu$ depends only on the ray $[\psi]\in\mathbb{P}(E_x)$:
$\mu(P\,|\,e^{i\theta}\psi)=\mu(P\,|\,\psi)$ for all $\theta\in\mathbb{R}$. This means that we assume the probability map depends only on the state’s ray, so multiplying $\psi$ by any global phase $e^{i\theta}$ does not change probabilities. We normalize the probability map so that for every state $\psi$, the~identity projector occurs with a probability of 1 and the zero projector with probability 0$\mu(\mathbf{1}\,|\,\psi)=1$ and $\mu(0\,|\,\psi)=0$. If~$\{P_k\}$ are mutually orthogonal ($P_jP_k=\delta_{jk}P_k$) and $P=\sum_k P_k$ converges, then
$\mu(P\,|\,\psi)=\sum_k \mu(P_k\,|\,\psi)$. 
If $P$ and $Q$ have the same range, then $\mu(P\,|\,\psi)=\mu(Q\,|\,\psi)$. That means there is additivity for pairwise orthogonal projectors $\{P_k\}$ with $P=\sum_k P_k$, 
$\mu(P\,|\,\psi)=\sum_k\mu(P_k\,|\,\psi)$. And~noncontextuality for subspaces if ${\rm Im}\,P={\rm Im}\,Q$ then $\mu(P\,|\,\psi)=\mu(Q\,|\,\psi)$. If~the projectors $\{P_k\}$ are mutually orthogonal  $P_jP_k=\delta_{jk}P_k$ and $P=\sum_k P_k$, with~the sum convergent, then the probability of $P$ given $\psi$ equals the sum of the individual probabilities, which is $\mu(P\,|\,\psi)=\sum_k \mu(P_k\,|\,\psi)$. If~two projectors $P$ and $Q$ have the same range, then they are assigned the same probability, $\mu(P\,|\,\psi)=\mu(Q\,|\,\psi)$.

For $\dim E_x\ge 3$, by~Gleason’s theorem, the~existence of a positive trace-one operator
$\rho_{[\psi]}$ on $E_x$ such that for every projector $P\in\mathcal{P}$:
\be
\mu(P\,|\,\psi)=\mathrm{Tr}(\rho_{[\psi]} P).
\ee
ray invariance and $U(1)$-equivariance of $E$ force $\rho_{[\psi]}$ to be a rank-one projector:
\(
\rho_{[\psi]}=\frac{|\psi\rangle\langle\psi|}{\langle\psi,\psi\rangle}.
\)
This shows:
\be
\mu(P\,|\,\psi)=\frac{\langle\psi,P\psi\rangle}{\langle\psi,\psi\rangle}.
\ee
Taking $P=\Pi_i$ gives the Born rule:
\be
\mu(\Pi_i\,|\,\psi)=\|\Pi_i\psi\|^2/\|\psi\|^2.
\ee

When $\dim E_x=2$, we replace A3 by $\sigma$-additivity for measurable fields of positive effects summing to $\mathbf{1}$ positive operator-valued measures (POVMs).
By the Busch--Gleason extension~\cite{BuschLahtiBook}, we again obtain
\(
\mu(E\,|\,\psi)=\mathrm{Tr}(\rho_{[\psi]} E)
\)
for all effects $E$; hence, we have the same Born rule for~projectors.

The preceding information is fiberwise. To~obtain spacetime probabilities for configuration-space localized outcomes, we must fix a Cauchy hypersurface $\Sigma\subset M$ with induced positive measure $\mathrm{d}\Sigma\sqrt{\gamma}$ from the real-slice metric.
For a Hermitian line subbundle $L\subset E$ such as a position or detector mode like a localized wave-packet subbundle defined by a smooth section basis, the~probability density is the fiber norm induced by $h$:
\be
\rho(x)=\frac{h_x(\Pi_L \psi_x,\Pi_L \psi_x)}{\int_\Sigma h_y(\psi_y,\psi_y)\,\mathrm{d}\Sigma_y\sqrt{\gamma(y)}}.
\ee
Gauge covariance under the $U(1)$ factor of the HUFT structure group is guaranteed because $h$ and $D$ are unitary,
$h(\psi,\psi)$ is $U(1)$-invariant, and~parallel transport preserves the norm. Here, $h(\psi,\psi)\equiv h_x(\psi_x,\psi_x)$ denotes the Hermitian fiber norm squared of $\psi$ at $x$; in~a local frame $(h_{i\bar j})$ with $\psi=\psi^i e_i$, we have $h(\psi,\psi)=\bar\psi^{\bar j}h_{i\bar j}\psi^i$. Thus, the~only $U(1)$-invariant quadratic functional compatible is the squared $h$-norm.

\noindent We let $E_x=E^S_x\otimes E^E_x$ describe system and environment fibers. For~a Schmidt state $|\psi\rangle=\sum_k \sqrt{p_k}\,|k\rangle_S\otimes|k\rangle_E$,
HUFT’s unitary fiber symmetries contain phase twirls
$U_S\otimes U_E^\dagger$ that leave $|\psi\rangle$ invariant called environment-assisted invariance~\cite{ZurekEnvariance}.
For equal-weight cases, $p_k$ equal force equiprobability on the system outcomes by symmetry, rational weights follow by refinement, and~continuity from (A3) yields
\(
\mu(\Pi_k\,|\,\psi)=p_k=\|\Pi_k\psi\|^2/\|\psi\|^2.
\)
Within HUFT, states are rays in a Hermitian vector bundle, and~measurements are Hermitian bundle endomorphisms.
Assuming ray invariance, additivity on orthogonal outcomes, and~locality or noncontextuality for subspaces, the~unique probability assignment is:
\be \Pr(a_i|\psi)=\frac{\langle\psi,\Pi_i\psi\rangle}{\langle\psi,\psi\rangle}=\frac{\|\Pi_i\psi\|_{h}^2}{\|\psi\|_{h}^2},
\ee
fiberwise at each spacetime point, and~its spacetime version is obtained by integrating the $h$-norm density over the appropriate hypersurface measure from the real-slice metric.
This is the Born rule in HUFT’s fiber-bundle~language.

The geometric setup described before holds, but~we also assume a ${\rm Spin}^c$ structure to derive the Dirac equation. We let $S\to M$ be the complex spinor bundle with Hermitian fiber metric
$h_S$ and Clifford map $c:T^*M\to{\rm End}(S)$, $c(\alpha)c(\beta)+c(\beta)c(\alpha)=2\,h^{\mu\nu}\alpha_\mu\beta_\nu\mathbf{1}$, meaning $c$ is a map $c:T^{\ast}M\to\mathrm{End}(S)$. For~any one-forms $\alpha,\beta$:
\be
c(\alpha)c(\beta)+c(\beta)c(\alpha)
=2\,h^{\mu\nu}\alpha_{\mu}\beta_{\nu}\,\mathbf{1}.
\ee
Equivalently, with~$c(dx^\mu)=\gamma^\mu$, this reads
$\{\gamma^\mu,\gamma^\nu\}=2\,h^{\mu\nu}\mathbf{1}$, as~this is the gamma-matrix anticommutation relation. For~the connection, we let $\nabla^{\rm LC}$ be the Levi--Civita connection of $h_{\mu\nu}$,
$\omega^{ab}{}_\mu$ its spin connection, and~$A_\mu\equiv A_\mu^{A} T_A + A^{(1)}_\mu$ the $G\times U(1)$ gauge potential.
The total unitary covariant derivative on sections of $S\otimes E_{\rm rep}$ is:
\be
D_\mu = \partial_\mu + \tfrac{1}{4}\,\omega^{ab}{}_\mu\gamma_{ab} + i\,A^{(1)}_\mu\,\mathbf{1}
          + i\,A_\mu^A T_A,
\quad\text{with}\quad \gamma_\mu \equiv c(e_\mu).
\ee
Compatibility means $D_\mu h_S=0$ and $[D_\mu,c(e_\nu)]=\Gamma^\rho_{\mu\nu}c(e_\rho)$; this is equivalent to metric or Clifford compatibility and ensures $D_\mu h_S=0$. We consider the Dirac Lagrangian density on the real slice~\cite{WaldGR, ParkerToms}:
\be
\mathcal{L}_D = \sqrt{|h|}\;\bar\psi\,(i\,\slashed{D}-m)\,\psi,
\qquad
\slashed{D}\equiv \gamma^\mu D_\mu,\quad
\bar\psi\equiv\psi^\dagger \gamma^0,
\ee
for $\psi\in\Gamma(S\otimes E_{\rm rep})$, this reads~as $\psi$ is a section of the tensor--product bundle $S\otimes E_{\rm rep}$, meaning at each spacetime point at $x\in M$:
\be
\psi(x)\in S_x \otimes (E_{\text{rep}})_x,
\ee
so $\psi$ is a spinor field from $S$ that also carries an internal gauge index from $E_{\rm rep}$. Here, $\Gamma(\cdot)$ denotes the space of the smooth sections. In~components, one can think of $\psi^{ai}(x)$ with spinor index $a$ and internal representation index $i$.
Varying with respect to \ $\bar\psi$ and using metric or Clifford compatibility gives:
\be
\delta_{\bar\psi}S_D \;=\;\int_M \sqrt{|h|}\;\delta\bar\psi\,(i\,\slashed{D}-m)\psi \;=\;0
\quad\Rightarrow\quad
(i\,\slashed{D}-m)\psi=0,
\ee
such as:
\be
i\,\gamma^\mu\Big(\partial_\mu + \tfrac{1}{4}\omega^{ab}{}_\mu\gamma_{ab} + i\,A^{(1)}_\mu + i\,A_\mu^A T_A\Big)\psi
\;-\; m\,\psi\;=\;0.
\ee
This means the Dirac equation is the precise covariant, unitary, Clifford-compatible section equation on the spinor bundle determined by HUFT’s geometry and gauge structure. On the complex slice, we would use the holomorphic or anti-holomorphic split, but~restriction to $M$ yields the~above.

Now we consider the one-loop effective action around slowly varying backgrounds. Let $\Phi$ denote the real components of the Higgs doublet $H$, gauge and ghost fields, and~fermions. The~quadratic fluctuation operator has a Laplace type on each associated~bundle:
\be
\Delta \;=\; -\,D^2 \;+\; \mathcal{E},
\qquad
D^2 \equiv h^{\mu\nu}D_\mu D_\nu,
\ee
where $\mathcal{E}\in\Gamma({\rm End}(\mathcal{V}))$ is an endomorphism built from background fields such as Higgs potential curvature, Yukawa endomorphisms, covariant curvatures of $D_\mu$, and~$\mathcal{V}$ is the vector bundle that the fluctuation field lives in.
The regulated one-loop effective action can be written via the heat kernel as~\cite{DeWitt1965,Vassilevich2003}:
\be
\Gamma^{(1)} \sim \tfrac{1}{2}\,{\rm STr}\,\log\Delta
\;\propto\;
\int_M \sqrt{|h|}\;\Big[
\frac{\Lambda^4}{(4\pi)^2}\,a_0
\;+\; \frac{\Lambda^2}{(4\pi)^2}\,a_1
\;+\;\cdots\Big],
\ee
with Seeley--DeWitt coefficients $a_k$ that are local, gauge- and diffeo-invariant fiber traces.
The quadratic divergence is controlled by:
\be
a_1 \;=\; {\rm STr}\,\mathcal{E},
\ee
a supertrace over all fluctuating species, bosons with $(+)$ sign, and fermions with $(-)$ sign, including~ghosts. Specializing in backgrounds where only $H$ is nonzero and slowly varying, $\mathcal{E}$ restricted to each species reduces at leading order to a mass-squared endomorphism that is affine in~$H^\dagger H$:
\be
\mathcal{E}\big|_{\text{species}}\;=\; m_0^2 \mathbf{1}\;+\; c_{\text{sp}}\,\lambda_{\text{sp}}\,(H^\dagger H)\,\mathbf{1}\;+\;\cdots,
\ee
with $c_{\text{sp}}$ a group-theory or counting factor and $\lambda_{\text{sp}}$ is the relevant coupling.
The coefficient of the induced operator $H^\dagger H$ in the quadratically divergent part of $\Gamma^{(1)}$ is proportional to the fiberwise supertrace:
\be
{\cal C}_{H^\dagger H}
\;\;=\;\;
{\rm STr}\,\big[\partial_{H^\dagger H}\mathcal{E}\big]
\;\;=\;\;
6\,\lambda
\;+\;\frac{9}{4}\,g^2
\;+\;\frac{3}{4}\,g'^2
\;-\;6\,y_t^2
\;+\;\cdots,
\ee
where the displayed terms are, respectively, the~Higgs self-coupling, ${\rm SU}(2)$ and ${\rm U}(1)_Y$ gauge couplings, and~the top Yukawa; the~ellipsis denotes smaller Yukawas and any additional fields in the chosen fiber~content.

The Veltman condition is treated as the fiber-geometric statement that this supertrace vanishes at some renormalization scale $\mu$~\cite{Veltman1981}:
\be
{\rm STr}\,\big[\partial_{H^\dagger H}\mathcal{E}\big](\mu)\;=\;0
\quad\Longleftrightarrow\quad
6\,\lambda(\mu)+\frac{9}{4}g^2(\mu)+\frac{3}{4}g'^2(\mu)-6\,y_t^2(\mu)+\cdots=0.
\ee
The scale $\mu$ where ${\rm STr}\,\partial_{H^\dagger H}\mathcal E(\mu)=0$ is scheme-dependent so the supertrace structure itself is the universal $a_1$ coefficient. The~interpretation of this is $a_1\propto{\rm STr}\,\mathcal{E}$ is a bundle trace of an endomorphism, so the sum rule is just the statement that the quadratic counterterm to $H^\dagger H$ disappears when the supertrace over all associated bundles, with~correct statistics and ghost structure,
is zero. In~HUFT with entire-function regulators, quadratic divergences are tamed, but~the same $a_1$ coefficient governs the finite threshold correction, replacing the naive $\Lambda^2$ piece, so the supertrace relation remains the geometric criterion for suppressing the Higgs mass~renormalization.

Using the same geometric structure as before, we
define a single holomorphic principal bundle~\cite{MoffatThompsonEmbedding2025}:
\be
\pi:\ \mathcal{Q}\to M,\qquad \mathrm{Struct}(\mathcal{Q})=\mathcal{H}:=\mathrm{SL}(2,\mathbb{C})_{\mathrm{Lor}}\times G,
\ee
equipped with a holomorphic connection $\mathcal{A}$ and curvature $\mathcal{F}:= \mathrm{d}\mathcal{A}+\mathcal{A}\wedge\mathcal{A}$. We use the Cartan block decomposition:
\be
\mathcal{A}=\omega\oplus A,\qquad \mathcal{F}=R\oplus F,
\ee
where $\omega$ is the complex spin connection, $R$ the complexified Riemann curvature, and~$A,F$ the internal gauge connection and curvature. We let $e$ denote the complex soldering form; on the real slice, $e$ induces $h=g_{(\mu\nu)}$. We now choose an $\mathrm{Ad}$-invariant, nondegenerate bilinear form on the Lie algebra:
\be
\langle X,Y\rangle_{\mathcal{H}}
:=\kappa_{\mathrm{Lor}}\operatorname{Tr}_{\mathrm{Lor}}(X_{\mathrm{Lor}}Y_{\mathrm{Lor}})
+\kappa_G\,\operatorname{Tr}_G(X_GY_G),
\qquad X= X_{\mathrm{Lor}}\oplus X_G,
\ee
with $\operatorname{Tr}$ the holomorphically normalized Killing forms on each factor.
We fix the ratio $\rho:=\kappa_{\mathrm{Lor}}/\kappa_G$ by the Chern--Weil matching condition:
\begin{equation}
\label{eq:CWfix}
\int_{M} \operatorname{Tr}_{\mathcal{H}}(\mathcal{F}\wedge\mathcal{F})
\ =\ 
\int_{M} \big(\kappa_{\mathrm{Lor}}\operatorname{Tr}_{\mathrm{Lor}}(R\wedge R)
+\kappa_G\operatorname{Tr}_G(F\wedge F)\big)\ \in\ 8\pi^2\,\mathbb{Z},
\end{equation}
compatible with the anomaly cancellation and boundary conditions specified below.
With this normalization, we define the one-coupling holomorphic action:
\begin{equation}
\label{eq:HUFTaction}
S_{\mathrm{HUFT}}
=\mathrm{Re} \int_{M} \Bigg[
\frac{1}{g_\star^{2}}\,
\langle \mathcal{F},\star \mathcal{F}\rangle_{\mathcal{H}}
\ +\ 
\lambda\,\langle B,\mathcal{F}\rangle_{\mathcal{H}}
\ +\
\mu\,\langle e\wedge e,\ R\rangle_{\mathrm{Lor}}
\ +\ S_{\mathrm{matter}}(\mathcal{A},e;\,\Psi)
\Bigg],
\end{equation}
where $B$ is an auxiliary holomorphic $\mathfrak{h}$-valued two-form and $\star$ is the complex Hodge operator. Varying \eqref{eq:HUFTaction} with respect to $B$ gives
\begin{equation}
\label{eq:Beq1}
\delta_B S:\qquad B=\frac{1}{\lambda}\,\star \mathcal{F}\quad\Rightarrow\quad B=\mathcal{F}\ \ \text{on the real slice},
\end{equation}
where the last statement follows after imposing reality or Hodge conditions on $(M,h)$.
Eliminating $B$ yields
\begin{equation}
\label{eq:reducedS}
S_{\mathrm{HUFT}}
=\mathrm{Re} \int_{M} \Big[
\frac{1}{g_\star^{2}}\,\langle \mathcal{F},\star \mathcal{F}\rangle_{\mathcal{H}}
+\mu\,\langle e\wedge e,\ R\rangle_{\mathrm{Lor}}
+S_{\mathrm{matter}}
\Big].
\end{equation}
Varying with respect to $\omega$ and $e$ gives, on~the real slice, metric compatibility and torsionlessness:
\begin{equation}
\nabla^{\mathrm{LC}}_\mu h_{\alpha\beta}=0,\qquad T(e,\omega)=0,
\end{equation}
and the Einstein equation sourced by the gauge--matter stress tensor. Variation with respect to $A$ gives the Yang--Mills equations with respect to $h$. Writing \eqref{eq:reducedS} on $(M,h)$:
\begin{equation}
\label{eq:EHYM}
S_{\mathrm{slice}}
=\int_{M} \Big[
\frac{1}{16\pi G_N}\,R(h)\,\mathrm{vol}_h
+\frac{1}{4g^2}\,\operatorname{Tr}_G(F_{\mu\nu}F^{\mu\nu})\,\mathrm{vol}_h
+\mathcal{L}_{\mathrm{matter}}
\Big],
\end{equation}
with:
\begin{equation}
\label{eq:ratio}
\frac{1}{16\pi G_N}:\frac{1}{4 g^2}\ =\ \mu\,\kappa_{\mathrm{Lor}}:\frac{1}{g_\star^2}\,\kappa_G
\ \ \Longrightarrow\ \
\frac{G_N}{g^2}\ =\ \frac{g_\star^2}{4\pi}\,\frac{\kappa_G}{\mu\,\kappa_{\mathrm{Lor}}}
\ =\ \frac{g_\star^2}{4\pi}\,\frac{1}{\mu\,\rho}.
\end{equation}
The relative gravity--gauge normalization is not a free dial as it is set by the geometric ratio $\rho=\kappa_{\mathrm{Lor}}/\kappa_G$ fixed in \eqref{eq:CWfix} and by the universal constants $(g_\star,\mu)$ entering \eqref{eq:HUFTaction}.

\textls[15]{The entire-function regulator $f(\Box/M_*^2)$ preserves the holomorphic Ward and Slavnov--Taylor} identities~\cite{Slavnov1972,Taylor1971,BRST1976}, yielding a common renormalization above $M_*$
\begin{equation}
\beta_{\mathcal{H}}(\mu>M_*)=0
\qquad\Rightarrow\qquad
g_{\mathrm{Lor}}(\mu)=g_G(\mu)=g_\star.
\end{equation}
Combined with \eqref{eq:ratio}, this gives a UV plateau with locked gravity--gauge normalization inherited by threshold matching to the~IR.

To show the internal gauge symmetries from our fibre bundle, we let $E\to M$ be a rank-$n$ holomorphic Hermitian vector bundle for matter, with~$c_1(E)=0$, a~nowhere-vanishing holomorphic volume form $\Omega_E\in H^0(M,\wedge^n E^*)$, and~assume $E$ is slope-stable so that the compatible connection is Hermitian--Yang--Mills

The dynamical HUFT action and field equations are formulated on the Lorentzian real slice $(M,h)$.
The appeal to slope stability and Donaldson--Uhlenbeck--Yau is not an additional hypothesis on $(M,h)$;
rather, it is a standard holomorphic-bundle criterion applied on an auxiliary compact Kähler base $(X,\omega)$
associated with the internal/matter bundle data. Restricting back to the real slice selects the same reduced unitary structure group such as $SU(n)$ when $c_1(E)=0$), while the Lorentzian dynamics proceed entirely with the usual gauge-natural constructions on $(M,h)$. Preservation of $(h_E,\Omega_E)$ reduces the internal structure group to $\mathrm{SU}(n)$, and~the internal connection has holonomy contained in~$\mathrm{SU}(n)$.

We demand a single adjoint reduction to the real-slice group, correct hypercharge quantization obtained from an integral pairing derived from \eqref{eq:CWfix}, and~chiral, anomaly-free matter for one family from associated bundles of $E$. Then, the minimal rank is $n=5$.

Now, for the mixed Ward identities to show the Diff $\leftrightarrow$ Gauge interlock, we
fix holomorphic gauge conditions for $\mathcal{A}$ and diffeomorphisms on $M$. To~preserve the gauge conditions, an~infinitesimal diffeomorphism $\delta_\xi$ must be accompanied by a compensating internal transformation $\delta_\epsilon$:
\begin{equation}
\delta_\xi \mathcal{A}+\delta_\epsilon \mathcal{A}=0\quad\text{in gauge},\qquad
\epsilon=\epsilon[\xi;\,\mathcal{A},e].
\end{equation}
The associated Ward identity for connected correlators $\langle \cdots \rangle$ is:
\begin{equation}
\partial_\mu \langle T^{\mu\nu}(x)\,\mathcal{O}(y)\rangle
=\alpha(\rho)\,\partial_\mu \langle J^{\mu}_{\mathrm{gauge}}(x)\,\mathcal{O}(y)\rangle+\cdots,
\qquad
\alpha(\rho)\propto \frac{\kappa_G}{\kappa_{\mathrm{Lor}}}=\rho^{-1},
\end{equation}
exhibiting an explicit cross-sector conservation relation with coefficient fixed by the geometric ratio $\rho$.

Finally, we will derive the Ehrenfest theorem from the unified geometry on the real slice $(M,h)$. We fix a global time function $t$ with Cauchy slices $\Sigma_t$,
unit normal $n^\mu$, induced metric $\gamma_{ij}$, and~measure $d\Sigma\,\sqrt{\gamma}$.
Let matter states be sections $\psi$ of the Hermitian bundle with $h$--inner product and
minimal coupling through the unitary covariant derivative:
\begin{equation}
D_\mu \;=\; \partial_\mu + \tfrac14 \omega^{ab}{}_\mu \gamma_{ab} + i\,A_\mu,
\qquad [D_\mu,D_\nu] \;=\; i\,F_{\mu\nu}.
\end{equation}
We use $\gamma^{a}$ ($a,b=0,1,2,3$) for the Dirac gamma matrices in a local Lorentz frame and:
\begin{equation}
\gamma^{ab}:=\tfrac12[\gamma^{a},\gamma^{b}]
\end{equation}
for their antisymmetrized products appearing in the spinor representation of $\mathfrak{spin}(1,3)$.
These objects are not related to the induced spatial metric $\gamma_{ij}$ ($i,j=1,2,3$) used below in
canonical or Schr\"odinger formulations.
When we foliate spacetime by spacelike hypersurfaces $\Sigma_t$, the~induced metric is
$\gamma_{\mu\nu}=h_{\mu\nu}+n_\mu n_\nu$ with unit normal $n^\mu$, and~$\gamma_{ij}$ denotes its pullback to
coordinates on $\Sigma_t$; we write $\sqrt{\gamma}:=\sqrt{\det(\gamma_{ij})}$ for the corresponding volume density. We fix a spacelike hypersurface $\Sigma_t$ with induced metric $\gamma_{ij}$ and volume form
$d\mu_{\Sigma_t}=\sqrt{\gamma}\,d^{3}x$.
The one-particle Hilbert space is $L^{2}(\Sigma_t,d\mu_{\Sigma_t})$ or its spinor analogue, with~inner product:
\begin{equation}
(\phi,\psi)_{\Sigma_t}:=\int_{\Sigma_t}\sqrt{\gamma}\,\phi^{*}\psi\,d^{3}x .
\end{equation}
A Schr\"odinger evolution is specified by an essentially self-adjoint Hamiltonian $\hat H$ on a dense domain,
so that $i\partial_t\psi=\hat H\psi$ and probability conservation is $(\psi,\psi)_{\Sigma_t}=\mathrm{const}$.

{We now will justify the appearance of the Schr\"odinger equation on the foliated real slice
from the geometric probability structure already defined above. For probabilities on $\Sigma_t$ and norm conservation, given a state $\psi$ on $\Sigma_t$, the Born density is the fiberwise Hermitian norm
integrated against the induced hypersurface measure $d\mu_{\Sigma_t}=\sqrt{\gamma}\,d^3x$,
so the total probability on $\Sigma_t$ is:}
\be
\|\psi\|^2_{\Sigma_t}:=(\psi,\psi)_{\Sigma_t}=\int_{\Sigma_t}\sqrt{\gamma}\,\psi^\dagger\psi\,d^3x.
\ee
We impose the physical requirement that total probability is independent of the chosen Cauchy slice:
\be
\frac{d}{dt}\|\psi\|^2_{\Sigma_t}=0.
\label{eq:norm_conservation_requirement}
\ee
This is the geometric content of probability conservation in the canonical picture. Next, identifying the Hilbert spaces at different $t$, because the inner product depends on the slice, we reduce to a fixed Hilbert space so we let $\Phi_t:\Sigma_0\to\Sigma_t$ be the diffeomorphism generated by the foliation flow such as the normal flow or equivalently, a choice of lapse/shift fixing an identification of points between slices. We write $J_t(x)$ for the Jacobian determinant relating the induced volume
forms:
\be
(\Phi_t)^*(\sqrt{\gamma(t)}\,d^3x)=J_t(x)\,\sqrt{\gamma(0)}\,d^3x.
\ee
Define an isometry $I_t:\mathcal{H}_t\to\mathcal{H}_0$ by the weighted pullback:
\be
(I_t\psi_t)(x):=J_t(x)^{1/2}\,\psi_t(\Phi_t(x)),
\label{eq:It_isometry_def}
\ee
so that for any $\phi_t,\psi_t\in\mathcal{H}_t$:
\begin{align}
(\,I_t\phi_t,\,I_t\psi_t\,)_{\Sigma_0}
&=\int_{\Sigma_0}\sqrt{\gamma(0)}\, (I_t\phi_t)^\dagger(I_t\psi_t)\,d^3x \\
&=\int_{\Sigma_0}\sqrt{\gamma(0)}\, J_t\,\phi_t(\Phi_t(x))^\dagger\psi_t(\Phi_t(x))\,d^3x \\
&=\int_{\Sigma_t}\sqrt{\gamma(t)}\,\phi_t^\dagger\psi_t\,d^3x
=(\phi_t,\psi_t)_{\Sigma_t}.
\end{align}
Hence $I_t$ is unitary, an inner-product preserving identification. Unitary time evolution and the one-parameter group come when we let $U_{t\leftarrow 0}:\mathcal{H}_0\to\mathcal{H}_t$ denote the physical evolution map
from $\Sigma_0$ to $\Sigma_t$, so that $\psi_t=U_{t\leftarrow 0}\psi_0$.
Probability conservation \eqref{eq:norm_conservation_requirement} implies:
\be
\|\psi_t\|_{\Sigma_t}=\|\psi_0\|_{\Sigma_0}
\quad \Longleftrightarrow \quad
(U_{t\leftarrow 0}\phi_0,\,U_{t\leftarrow 0}\psi_0)_{\Sigma_t}=(\phi_0,\psi_0)_{\Sigma_0}.
\ee
Equivalently, the map $\widetilde U(t):=I_t\circ U_{t\leftarrow 0}:\mathcal{H}_0\to\mathcal{H}_0$
is unitary:
\be
(\widetilde U(t)\phi_0,\widetilde U(t)\psi_0)_{\Sigma_0}=(\phi_0,\psi_0)_{\Sigma_0}.
\label{eq:U_unitary}
\ee
We assume the standard composition and continuity properties of time evolution:
\be
\widetilde U(0)=\mathbf{1},\qquad
\widetilde U(t+s)=\widetilde U(t)\widetilde U(s),\qquad
\widetilde U(t)\ \text{strongly continuous in }t.
\label{eq:U_group_properties}
\ee
Then $\{\widetilde U(t)\}_{t\in\mathbb{R}}$ is a strongly continuous one-parameter unitary group on
$\mathcal{H}_0$. The existence of a self-adjoint generator and the Schr\"odinger equation. By Stone's theorem or more precisely the Stone–von Neumann theorem, there exists a densely-defined self-adjoint operator $\widehat H$ on $\mathcal{H}_0$
such that:
\be
\widetilde U(t)=\exp\!\left(-\frac{i}{\hbar}\,t\,\widehat H\right).
\label{eq:stone_unitary_group}
\ee
Differentiating \eqref{eq:stone_unitary_group} in the strong sense yields:
\be
\frac{d}{dt}\widetilde U(t)=-\frac{i}{\hbar}\,\widehat H\,\widetilde U(t).
\label{eq:U_generator_ode}
\ee
\textls[-25]{Applying $\widetilde U(t)$ to the initial state $\psi_0\in\mathrm{Dom}(\widehat H)$ defines
$\widetilde\psi(t):=\widetilde U(t)\psi_0\in\mathcal{H}_0$, and \eqref{eq:U_generator_ode} gives:}
\be
i\hbar\,\partial_t\widetilde\psi(t)=\widehat H\,\widetilde\psi(t).
\label{eq:schrodinger_fixed_hilbert}
\ee
Returning to the time-dependent slice using $\psi_t=I_t^{-1}\widetilde\psi(t)$ gives the canonical
Schr\"odinger form on $\Sigma_t$:
\be
i\hbar\,\partial_t\psi_t=\widehat H_t\,\psi_t,
\qquad
\widehat H_t:=I_t^{-1}\widehat H\,I_t,
\label{eq:schrodinger_on_slice}
\ee
with $\widehat H_t$ self-adjoint with respect to $(\cdot,\cdot)_{\Sigma_t}$.
Thus, once the geometric probability assignment and norm conservation are imposed, the existence
of a Schr\"odinger generator is not an additional assumption but follows from unitarity.

The real-slice geometry provides the induced spatial metric $\gamma_{ij}$ and the unitary covariant derivative
$D_\mu$ including spin and internal gauge pieces. We denote by $D_i$ the pullback of $D_\mu$ to $\Sigma_t$
and define the spatial kinetic momentum operator:
\be
\widehat\Pi_i:=-i\hbar\,D_i.
\ee
The natural gauge- and diffeomorphism-covariant Laplace--Beltrami operator on $\Sigma_t$ is:
\be
\Delta_{\gamma,D}\psi
:=\frac{1}{\sqrt{\gamma}}\,D_i\!\left(\sqrt{\gamma}\,\gamma^{ij}D_j\psi\right),
\label{eq:covariant_laplace_beltrami}
\ee
which is symmetric on $L^2(\Sigma_t,d\mu_{\Sigma_t})$ under standard falloff/boundary conditions.
For a spin-0 Schr\"odinger field with real scalar potential $V$, the minimally coupled \mbox{Hamiltonian is:}
\be
\widehat H_t
=
-\frac{\hbar^2}{2m}\,\Delta_{\gamma,D}
+V,
\label{eq:geometric_schrodinger_hamiltonian}
\ee
and for charged matter one includes the appropriate $U(1)$ or nonabelian temporal connection component
inside $D_t$ (equivalently as a $qA_0$ term in $\widehat H_t$ depending on conventions).
More generally, the Pauli/Dirac Hamiltonians are obtained by replacing \eqref{eq:geometric_schrodinger_hamiltonian}
with the corresponding first-order operators built from $(h_{\mu\nu},D_\mu)$, and the above unitary-generator
argument applies verbatim.

Under these standard hypotheses self-adjointness and suitable boundary conditions, the~Heisenberg~identity:
\begin{equation}
\frac{d}{dt}\langle\hat A\rangle
=\frac{i}{\hbar}\langle[\hat H,\hat A]\rangle+\left\langle\frac{\partial \hat A}{\partial t}\right\rangle
\end{equation}
follows by differentiating $\langle\hat A\rangle$ and substituting the Schr\"odinger equation. We define the kinetic momentum operator $\hat\Pi_\mu:=-i\hbar D_\mu$, the~position operator $\hat X^\mu$, multiplication by the coordinate function on $\Sigma_t$, and~for any bundle endomorphism observable $\hat O$ define the expectation value on $\Sigma_t$ by:
\begin{equation}
\label{eq:expectation}
\langle \hat O\rangle_t
\;:=\;
\frac{\displaystyle\int_{\Sigma_t}\!\sqrt{\gamma}\;\psi^\dagger \hat O\,\psi}
     {\displaystyle\int_{\Sigma_t}\!\sqrt{\gamma}\;\psi^\dagger\psi}.
\end{equation}
We take dynamics to be generated by the minimally coupled covariant Hamiltonian $\hat H$, either Dirac, Pauli, or~Schr\"odinger built from $h$ and $D_\mu$, so that $i\hbar\,\partial_t\psi=\hat H\psi$ and the
probability current is~conserved.

We assume standard falloff, so boundary terms at spatial infinity vanish, and~take $V$ a real scalar section such as a gauge singlet potential. Then, the expectation values of position and kinetic momentum obey:
\begin{align}
\label{eq:ehrenfest_pos}
\frac{D}{dt}\,\big\langle \hat X^\mu\big\rangle
&= \frac{1}{m}\,\big\langle \hat\Pi^\mu\big\rangle,\\[4pt]
\label{eq:ehrenfest_mom}
\frac{D}{dt}\,\big\langle \hat\Pi_\mu\big\rangle
&= -\,\big\langle \nabla_\mu V\big\rangle
   + \big\langle q\,F_{\mu\nu}\,\dot X^\nu\big\rangle
   - \big\langle \Gamma^\rho_{\mu\nu}\,\hat\Pi_\rho\,\dot X^\nu\big\rangle,
\end{align}
where $\Gamma^\rho_{\mu\nu}$ is the Levi--Civita connection of $h$, the~overdot denotes contraction
with $n^\mu$ as the foliation flow, and~$F_{\mu\nu}$ is the gauge curvature which equals the antisymmetric piece $B_{\mu\nu}$ of the Hermitian metric on the real slice ($B=F$). The~derivative $D/dt$ is the
Levi--Civita covariant time derivative along the foliation, so that measure and connection effects are~included.

Conservation of the probability current gives the covariant continuity equation
$\partial_t(\sqrt{\gamma}\,\psi^\dagger\psi)+\nabla_i(\sqrt{\gamma}\,j^i)=0$, so boundary terms
from $\partial_t\sqrt{\gamma}$ and $\nabla_i$ cancel in the time derivative of \eqref{eq:expectation}.
Thus, we obtain the Heisenberg identity in covariant form:
\begin{equation}
\label{eq:heisenberg}
\frac{D}{dt}\langle \hat O\rangle
=\frac{i}{\hbar}\,\langle[\hat H,\hat O]\rangle
+\langle \partial_t \hat O\rangle,
\end{equation}
with the Levi--Civita correction precisely reproducing the $\Gamma$--terms~below. For $\hat O=\hat X^\mu$, minimal coupling implies
$[\hat \Pi_\alpha,\hat X^\mu]=-i\hbar\,\delta^\mu{}_\alpha$, and~for the kinetic Hamiltonian
$\hat H_{\rm kin}=\tfrac{1}{2m}\,h^{\alpha\beta}\hat\Pi_\alpha\hat\Pi_\beta$ we obtain
\be
\frac{i}{\hbar}[\hat H_{\rm kin},\hat X^\mu]
= \frac{1}{2m}\,h^{\alpha\beta}\Big(
\hat\Pi_\alpha[\hat\Pi_\beta,\hat X^\mu]+[\hat\Pi_\alpha,\hat X^\mu]\hat\Pi_\beta\Big)
= \frac{1}{m}\,\hat\Pi^\mu,
\ee
yielding \eqref{eq:ehrenfest_pos} after taking expectation values and accounting for the foliation, with no
explicit time dependence of $\hat X^\mu$.

For $\hat O=\hat\Pi_\mu$, we use curvature and metric compatibility and define the kinetic momentum operator:
\be
\hat\Pi_\mu=-i\hbar\,D_\mu,
\ee 
with the kinetic Hamiltonian:
\begin{equation}
    \hat{H}_{\rm kin}=\frac{1}{2m}h^{\alpha\beta}\hat{\Pi}_\alpha\hat{\Pi}_\beta.
\end{equation}
Curvature shows up as a commutator, by~definition of the gauge curvature:
\be
[\hat\Pi_\mu,\hat\Pi_\nu]=-i\hbar\,F_{\mu\nu}.
\ee 
With the metric compatibility: 
\be
\nabla_\lambda h_{\mu\nu}=0,
\ee
meaning $h^{\alpha\beta}$ is covariantly constant, so it commutes with $\hat{\Pi}_\mu$ so there are no extra terms from derivatives of the metric when you commute things.
Then the Heisenberg force operator~follows:
\be
\frac{i}{\hbar}[\hat H_{\rm kin},\hat\Pi_\mu]
= \frac{1}{2m}\,h^{\alpha\beta}
\Big(\hat\Pi_\alpha[\hat\Pi_\beta,\hat\Pi_\mu]+[\hat\Pi_\alpha,\hat\Pi_\mu]\hat\Pi_\beta\Big)
= \frac{1}{m}\,h^{\alpha\beta}\,\Big(\hat\Pi_\alpha F_{\beta\mu}+F_{\alpha\mu}\hat\Pi_\beta\Big),
\ee
the right-hand side is the symmetrized product of momentum with field strength, so the operator is Hermitian. Its symmetric part gives the Lorentz-force term $q\,F_{\mu\nu}\,\dot X^\nu$ in expectation values.
The scalar potential contributes $(i/\hbar)[V,\hat\Pi_\mu]=-\nabla_\mu V$.
Finally, when transporting the expectation value with the time-dependent volume form and frame,
the Levi--Civita piece from $D/dt$ supplies $-\langle \Gamma^\rho_{\mu\nu}\,\hat\Pi_\rho\,\dot X^\nu\rangle$,
completing \eqref{eq:ehrenfest_mom}. Since on the real slice, $B_{\mu\nu}=F_{\mu\nu}$, the~gauge-force term is
literally the antisymmetric part of the Hermitian~metric.

In a local inertial frame at a point $\Gamma=0$, the relations are reduced by taking the expectation values and writing:
\begin{equation}
\hat{\Pi}^\nu:=h^{\nu\alpha}\hat{\Pi}_\alpha,
\end{equation}
and:
\begin{equation}
    \hat{v}^\nu:=\frac{\hat{\Pi}^\nu}{m},
\end{equation}
gives the Lorentz-force term in Ehrenfest form:
\begin{equation}
    \frac{d}{dt}\langle\hat{\Pi}_\mu\rangle=\left\langle\frac{i}{\hbar}[\hat{H}_{\rm kin}, \hat{H}_\mu]\right\rangle, \to q\langle F_{\mu\nu}\hat{v}^\nu\rangle,
\end{equation}
in curved space, we add the separate Levi--Civita piece but note that when you take the time derivative of the momentum expectation, you must use the covariant time derivative:
\begin{equation}
\frac{D}{dt}\langle\hat{\Pi}_\mu\rangle = \frac{d}{dt}\langle\hat{\Pi}_\mu\rangle -\langle \Gamma_{\mu\nu}^\rho \hat{\Pi}_\rho \dot{X}^\nu \rangle,
\end{equation}
When transporting the expectation value, that is the gravitational part. That extra term is what we call the Levi--Civita piece. It is the correction needed so that the equation of motion is tensorial coordinate-independent. Without~it, you would be differentiating components as if the basis were fixed, flat space. With~it, the~quantum expectation values reduce to the geodesic with Lorentz force in the classical~limit.

We find the Ehrenfest relations:
\begin{equation}
\frac{d}{dt}\langle \hat X^\mu\rangle=\frac{1}{m}\langle \hat\Pi^\mu\rangle,
\qquad
\frac{d}{dt}\langle \hat\Pi_\mu\rangle
= -\langle \nabla_\mu V\rangle + \langle q\,F_{\mu\nu}\,\dot X^\nu\rangle,
\end{equation}
geodesic drift plus the Lorentz force in~expectation.

On a static slice of flat space with coordinates $(t,\mathbf x)$, $\Gamma=0$ and $h_{\mu\nu}\to\eta_{\mu\nu}$.
Writing $\hat{\boldsymbol\pi}=-i\hbar\,\boldsymbol{\nabla}-q\,\mathbf A$ and $\hat{\mathbf v}=\hat{\boldsymbol\pi}/m$:
\begin{equation}
\frac{d}{dt}\langle \hat{\mathbf x}\rangle=\frac{1}{m}\langle \hat{\boldsymbol\pi}\rangle,\qquad
\frac{d}{dt}\langle \hat{\boldsymbol\pi}\rangle
= -\langle \boldsymbol{\nabla} V\rangle
+ q\,\big\langle \mathbf E + \hat{\mathbf v}\times \mathbf B\big\rangle,
\end{equation}
which collapses to the textbook Ehrenfest theorem when $\mathbf B=0$ and $V=V(\mathbf x)$.

Our observables are bundle endomorphisms compatible with the unitary structure, so this excludes non-covariant ad~hoc couplings and ensures the identities above are representation-independent. Gravity enters only through
$h$ via $\Gamma$, while gauge forces enter through $F$ but both arise from the same unified connection in the holomorphic theory, so the quantum to classical correspondence is an internal consequence of the geometry, not an
additional~postulate.

\section{Conclusions}
\label{sec:conclusion}


We have presented a symmetry–first unification in which gravity and gauge interactions arise from a single geometric framework, a~product bundle with one master connection $\mathcal{A}=(\omega,A)$, one curvature $\mathcal{F}=(R,F)$, and~a
Hermitian field $g=h+iB$ on the complexified spacetime. From~a single $\mathrm{Diff}(M)\times G$–invariant action,
variation reproduces the Einstein and Yang--Mills equations together with their paired Bianchi identities,
realizing the Einstein--Noether program of deriving dynamics and identities from one invariant~structure.

Within this unified kinematic and variational structure, there are no additional diffeomorphism- and gauge-natural, unitary, second-order couplings in four dimensions beyond Einstein--Hilbert, Yang--Mills, and~minimal matter terms. This makes the SM and GR content a consequence of symmetry and naturalness rather than an arbitrary choice, and~the Hermitian packaging renders this uniqueness geometrically and explicitly through $g_{[\mu\nu]}=F_{\mu\nu}$.

Quantum kinematics and measurement descend from the same data. The~states are rays in a Hermitian bundle with unitary
connection $D$, and~the unique probability assignment compatible with our axioms is the Born rule. Fiberwise, its
spacetime version follows by integrating the $h$-norm on Cauchy slices with the induced measure $\sqrt{\gamma}\,d\Sigma$.
This ties probabilities to the same real-slice geometry that governs~dynamics.

On the perturbative side, implementing entire-function regulators at covariant kinetic operators and vertices yields a
UV-finite framework, whose IR limit reproduces local EYM and matter. The~unified Noether-II content includes covariant
conservation laws and Slavnov--Taylor identities, ensuring consistency of the regulated~theory.

Taken together, these results realize a form of geometric unification in which parallel transport, curvature, conserved
currents, and~measurement all arise from the same $(h,A)$ data, with~$B=F$ eliminating superfluous antisymmetric
gravitational~modes.

Several targeted directions can be followed. A~group or metric locking principle, either geometric or topological, is needed to fix the relative normalization and uniquely determine the internal group, elevating our
shared-structure unification toward a strict GUT selection. On~the quantum–classical interface, the~covariant Ehrenfest lemma, linking the time evolution of expectation values in quantum theory to the classical equations of motion, shows that geodesic drift and Lorentz force emerge in expectation directly from the unified geometry, extending this to open-system dynamics and decoherence on curved~backgrounds.

\vspace{6pt}
\authorcontributions{{Conceptualization, E.J.T. and J.W.M.; methodology, E.J.T.; investigation, E.J.T. and J.W.M.; writing---original draft preparation, E.J.T.; writing---review and editing, E.J.T. and J.W.M.; supervision, J.W.M. All authors have read and agreed to the published version of the manuscript.}}

\funding{{This research received no external funding.}}

\dataavailability{No new data were created or analyzed in this study. Data sharing is not applicable.}

\acknowledgments{We thank 
 Martin Green and Arvin Kouroshnia for helpful discussions. We thank Gabriela Secara for visualization.  Research at the Perimeter Institute for Theoretical Physics is supported by the Government of Canada through Industry Canada and by the Province of Ontario through the Ministry of Research and Innovation~(MRI).} 

\conflictsofinterest{{The authors declare no conflicts of interest.}}
\clearpage 
\begin{adjustwidth}{-\extralength}{0cm}

\reftitle{References}

\PublishersNote{}
\end{adjustwidth}

\begin{thebibliography}{999}
\bibitem{Einstein1945}
{Einstein, A.} 
\textls[-25]{A Generalization of the Relativistic Theory of Gravitation.
\emph{Ann. Math.} \textbf{1945}, {\textit{46}}, 578--584. [\href{http://dx.doi.org/10.2307/1969197}{CrossRef}]}
\bibitem{NoetherTrans}
Noether, E. Invariant Variational Problems.
{English translation by M.~A.~Tavel.}  \textit{arXiv} \textbf{1971}, arXiv:physics/0503066.
\bibitem{WaldGR}
Wald, R.M.  \emph{General Relativity}; University of Chicago Press: {Chicago, IL, USA}, 1984.
\bibitem{YangMills1954}
Yang, C.N.; Mills, R.L.
Conservation of Isotopic Spin and Isotopic Gauge Invariance.
\emph{Phys.\ Rev.} \textbf{1954}, {\textit{96}}, 191--195. [\href{http://dx.doi.org/10.1103/PhysRev.96.191}{CrossRef}]
\bibitem{KobayashiNomizu1}
Kobayashi, S.; Nomizu, K.
\emph{Foundations of Differential Geometry};  Wiley:  {Hoboken, NJ, USA,}   1963; Volume~1.
\bibitem{Nakahara}
Nakahara, M.
\emph{Geometry, Topology and Physics}, 2nd ed.;  Taylor \& Francis: {Hoboken, NJ, USA,} 2003.
\bibitem{Trautman1979}
Trautman, A.   Fiber bundles, gauge fields, and gravitation. In
\emph{General Relativity and Gravitation}; Held, A., Ed.;  Plenum: {New York, NY, USA}, 1980; Volume~1, pp. 287--308.
\bibitem{Slavnov1972}
Slavnov, A.A. Ward Identities in Gauge Theories.
\emph{Theor.\ Math.\ Phys.} \textbf{1972}, {\textit{10}},  99--107. [\href{http://dx.doi.org/10.1007/BF01090719}{CrossRef}]
\bibitem{Taylor1971}
Taylor, J.C. Ward Identities and Charge Renormalization of the Yang–Mills Field.
\emph{Nucl.\ Phys.\ B} \textbf{1971}, \textit{33},  436--444. [\href{http://dx.doi.org/10.1016/0550-3213(71)90297-5}{CrossRef}]
\bibitem{BRST1976}
Becchi, C.; Rouet, A.; Stora, R. Renormalization of the abelian Higgs‐Kibble model.
\emph{Commun.\ Math.\ Phys.} \textbf{1975}, \textit{42}, 127--{162}. [\href{http://dx.doi.org/10.1007/BF01614158}{CrossRef}]
\bibitem{GeorgiGlashow1974}
Georgi, H.; Glashow, S.L.
Unity of All Elementary Particle Forces.
\emph{Phys.\ Rev.\ Lett.} \textbf{1974}, {\textit 32}, 438--441. [\href{http://dx.doi.org/10.1103/PhysRevLett.32.438}{CrossRef}]
\bibitem{Slansky1981}
Slansky, R. Group Theory for Unified Model Building.
\emph{Phys.\ Rept.} \textbf{1981}, \textit{79}, 1--128. [\href{http://dx.doi.org/10.1016/0370-1573(81)90092-2}{CrossRef}]
\bibitem{DeWitt1965}
DeWitt, B.S.   \emph{Dynamical Theory of Groups and Fields}; Gordon and Breach: {London, UK}, 1965.
\bibitem{Vassilevich2003}
Vassilevich, D.V. Heat kernel expansion: User's manual.
\emph{Phys.\ Rep.} \textbf{2003}, \textit{388}, 279--{360}. [\href{http://dx.doi.org/10.1016/j.physrep.2003.09.002}{CrossRef}]
\bibitem{Bardeen1969}
Bardeen, W.A. Anomalous Ward Identities in Spinor Field Theories.
\emph{Phys.\ Rev.} \textbf{1969}, \textit{184}, 1848--1857. [\href{http://dx.doi.org/10.1103/PhysRev.184.1848}{CrossRef}]
\bibitem{SusskindHologram1995}
Susskind, L.
The World as a Hologram.
\emph{J.\ Math.\ Phys.} \textbf{1995}, \textit{36}, 6377--{6396}. [\href{http://dx.doi.org/10.1063/1.531249}{CrossRef}]
\bibitem{SusskindLandscape2003}
Susskind, L.
The Anthropic Landscape of String Theory. \textit{arXiv} \textbf{2003},
arXiv:hep-th/0302219. [\href{http://dx.doi.org/10.48550/arXiv.hep-th/0302219}{CrossRef}]
\bibitem{Witten1995}
Witten, E.
String Theory Dynamics in Various Dimensions.
\emph{Nucl.\ Phys.\ B} \textbf{1995}, \textit{443}, 85--{126}. [\href{http://dx.doi.org/10.1016/0550-3213(95)00158-O}{CrossRef}]
\bibitem{WittenAdS1998}
\textls[-15]{Witten, E. Anti De Sitter Space and Holography.
\emph{Adv.\ Theor.\ Math.\ Phys.} \textbf{1998}, \textit{2}, 253--{291}. [\href{http://dx.doi.org/10.4310/ATMP.1998.v2.n2.a2}{CrossRef}]}
\bibitem{PenroseTwistor1976}
Penrose, R.
Nonlinear Gravitons and Curved Twistor Theory.
\emph{Gen.\ Rel.\ Grav.} \textbf{1976}, \textit{7}, 31--52. [\href{http://dx.doi.org/10.1007/BF00762011}{CrossRef}]
\bibitem{KakuKikkawaTrees1974}
Kaku, M.; Kikkawa, K.
Field Theory of Relativistic Strings. I. Trees.
\emph{Phys.\ Rev.\ D} \textbf{1974}, \textit{10}, 1110--1133. [\href{http://dx.doi.org/10.1103/PhysRevD.10.1110}{CrossRef}]
\bibitem{KakuKikkawaLoops1974}
Kaku, M.; Kikkawa, K.
Field Theory of Relativistic Strings. II. Loops and Pomerons.
\emph{Phys.\ Rev.\ D} \textbf{1974}, \textit{10}, 1823--1843. [\href{http://dx.doi.org/10.1103/PhysRevD.10.1823}{CrossRef}]
\bibitem{KakuClosedSFT1988}
Kaku, M. Geometric Derivation of String Field Theory from First Principles: Closed Strings and Modular Invariance.
\emph{Phys.\ Rev.\ D} \textbf{1988}, \textit{38}, 3052--3060. [\href{http://dx.doi.org/10.1103/PhysRevD.38.3052}{CrossRef}]
\bibitem{MoffatThompsonHUFT2025}
Moffat, J.W.; Thompson, E.J. Holomorphic Unified Field Theory of Gravity and the Standard Model.  \textit{Eur. Phys. J. C} \textbf{2025},  \textit{85}, {1157}. [\href{http://dx.doi.org/10.1140/epjc/s10052-025-14907-2}{CrossRef}]
\bibitem{MoffatThompsonFiniteNonlocal2025}
Moffat, J.W.; Thompson, E.J.
Finite Nonlocal Holomorphic Unified Quantum Field Theory. \textit{arXiv}  \textbf{2025},
arXiv:2507.14203.
\bibitem{MoffatThompsonMassSpectrum2025}
Moffat, J.W.; Thompson, E.J.
On the Standard Model Mass Spectrum and Interactions in the Holomorphic Unified Field Theory.  \textit{arXiv}  \textbf{2025}, arXiv:2508.02747.
\bibitem{ThompsonTopQuark2025}
Thompson, E.J.
Top Quark Bound States in Finite and Holomorphic Quantum Field Theories.  \textit{arXiv}  \textbf{2025}, arXiv:2507.16831v2.
\bibitem{MT:ReplyToCline}
Moffat, J.W.; Thompson, E.J.
Comment on a ``Comment on ``Standard Model Mass Spectrum and Interactions in the Holomorphic Unified Field Theory''''. \textit{arXiv}  \textbf{2025}, arXiv:2508.08510. [\href{http://dx.doi.org/10.48550/arXiv.2508.08510}{CrossRef}]
\bibitem{MT:GI2025}
Moffat, J.W.; Thompson, E.J.
Gauge-Invariant Entire-Function Regulators and UV Finiteness in Non-Local Quantum Field Theory. \textit{arXiv}  \textbf{2025},  arXiv:2511.11756.
\bibitem{MT:AdSdS2025}
Moffat, J.W.; Thompson, E.J.
On the Complexified Spacetime Manifold Mapping of AdS to dS. \textit{arXiv}  \textbf{2025},
arXiv:2511.11658.
\bibitem{T:DarkMatter2025}
Thompson, E.J.
$\mathbb Z_2$-Stable Dark Matter via Broken $\text{SU}(5)$ Gauge Bosons. \textit{arXiv}  \textbf{2025},
arXiv:2511.19462v1.
\bibitem{UhlenbeckYau1986}
Uhlenbeck, K.; Yau, S.-T. On the existence of Hermitian–Yang–Mills connections in stable vector bundles.
\emph{Comm.\ Pure Appl.\ Math.} \textbf{1986}, \textit{39}, S257--S293. [\href{http://dx.doi.org/10.1002/cpa.3160390714}{CrossRef}]
\bibitem{Donaldson1985}
Donaldson, S.K. Anti–self-dual Yang–Mills connections over complex algebraic surfaces and stable vector bundles.
\emph{Proc.\ Lond.\ Math.\ Soc.} \textbf{1985}, \textit{50}, 1--26. [\href{http://dx.doi.org/10.1112/plms/s3-50.1.1}{CrossRef}]
\bibitem{Music}
Burtscher, A.
\textit{Introduction to Riemannian Geometry.
Course Module NWI-WM310 (Section 2.2.1—“Raising and Lowering Indices”)}; Fall semester 2024/25;
Radboud University Nijmegen: {Nijmegen, The Netherlands}, 21 February 2025.
\bibitem{Curvature}
Lee, J.M.
\textit{Riemannian Manifolds: An Introduction to Curvature (Section 3 ``Definitions and Examples of Riemannian Metrics'')};  {Springer: Berlin/Heidelberg, Germany,} 1977.
\bibitem{DadhichPons2012}
Dadhich, N.; Pons, J.M.
On the equivalence of the Einstein--Hilbert and the Einstein--Palatini formulations of general relativity for an arbitrary connection.
\textit{Gen. Relativ. Gravit.} \textbf{2012}, \textit{44}, 2337--2352. [\href{http://dx.doi.org/10.1007/s10714-012-1393-9}{CrossRef}]
\bibitem{GibbonsHawking1977}
Gibbons, G.W.; Hawking, S.W.
\newblock  Action integrals and partition functions in quantum gravity.
\newblock {\em Phys. Rev. D} \textbf{1977}, \textit{15}, 2752. [\href{http://dx.doi.org/10.1103/PhysRevD.15.2752}{CrossRef}]
\bibitem{York1972}
York, J.W.,  Jr.
\newblock  Role of conformal three-geometry in the dynamics of gravitation.
\newblock {\em Phys. Rev. Lett.} \textbf{1972}, {\textit{28}}, 1082. [\href{http://dx.doi.org/10.1103/PhysRevLett.28.1082}{CrossRef}]
\bibitem{HehlEtAl1976}
Hehl, F.W.; von der Heyde, P.; Kerlick, G.D.; Nester, J.M.
General relativity with spin and torsion: Foundations and prospects.
\textit{Rev. Mod. Phys.} \textbf{1976}, \textit{48}, 393--416. [\href{http://dx.doi.org/10.1103/RevModPhys.48.393}{CrossRef}]
\bibitem{Holst1996}
Holst, S.
Barbero's Hamiltonian derived from a generalized Hilbert--Palatini action.
\textit{Phys. Rev. D} \textbf{1996}, \textit{53}, 5966--5969. [\href{http://dx.doi.org/10.1103/PhysRevD.53.5966}{CrossRef}] [\href{http://www.ncbi.nlm.nih.gov/pubmed/10019884}{PubMed}]
\bibitem{Palatini1919}
Palatini, A.
Deduzione invariantiva delle equazioni gravitazionali dal principio di Hamilton.
\textit{Rendiconti del Circolo Matematico di Palermo} \textbf{1919}, \textit{43}, 203--212. [\href{http://dx.doi.org/10.1007/BF03014670}{CrossRef}]
\bibitem{Gleason1957}
Gleason, A.M. Measures on the Closed Subspaces of a Hilbert Space.
\emph{J.\ Math.\ Mech.} \textbf{1957}, \textit{6}, 885--893. [\href{http://dx.doi.org/10.1512/iumj.1957.6.56050}{CrossRef}]
\bibitem{BuschLahtiBook}
Busch, P.; Lahti, P.; Mittelstaedt, P.
\emph{The Quantum Theory of Measurement}, 2nd ed.; Springer:  {Berlin/Heidelberg, Germany,} 1996.
\bibitem{ZurekEnvariance}
Zurek, W.H. Environment-Assisted Invariance, Entanglement, and Probabilities in Quantum Physics.
\emph{Phys.\ Rev.\ Lett.} \textbf{2003}, \textit{90}, 120404. [\href{http://dx.doi.org/10.1103/PhysRevLett.90.120404}{CrossRef}]
\bibitem{Veltman1981}
Veltman, M.J.G.
The Infrared–Ultraviolet Connection.
\emph{Acta Phys.\ Polon.\ B} \textbf{1981}, \textit{12}, 437--457.
\bibitem{ParkerToms}
Parker, L.; Toms, D.  \emph{Quantum Field Theory in Curved Spacetime};
Cambridge University Press: {Cambridge, UK}, 2009.
\bibitem{MoffatThompsonEmbedding2025}
Moffat, J.W.; Thompson, E.J.
Embedding $\mathrm{SL}(2,\mathbb{C})/\mathbb{Z}_2$ in Complex Riemannian Geometry. \textit{arXiv} \textbf{2025}, arXiv:2506.19158.
\end{thebibliography}
\end{document}